\documentclass[fleqn,10pt]{wlscirep}
\usepackage[T1]{fontenc}
\usepackage{subcaption} 
\usepackage{authblk}    
\usepackage{amsmath,amsfonts,amssymb}
\usepackage{mathptmx}   
\usepackage{float}
\usepackage{helvet}
\usepackage{courier}
\usepackage{ifpdf}
\usepackage{graphicx,xcolor}
\usepackage{multirow}
\usepackage{longtable}
\usepackage{tabularx}
\usepackage{booktabs}
\usepackage{url}
\usepackage{titletoc}
\usepackage{graphicx}

\title{Causal Links Between Anthropogenic Emissions and Air Pollution Dynamics in Delhi}

\author[1,*]{Sourish Das}
\author[2]{Sudeep Shukla}
\author[3]{Alka Yadav}
\author[4,*]{Anirban Chakraborti}
\affil[1]{Department of Mathematics, Chennai Mathematical Institute, Chennai-603103, India}
\affil[2]{AI 4 Water LTD, Orpington, BR6 9QX United Kingdom}
\affil[3]{Department of Physics, Bennett University, Greater Noida - 201310, India}
\affil[4]{School of Computational and Integrative Sciences, Jawaharlal Nehru University, New Delhi-110067, India}
\affil[*]{Corresponding authors: sourish@cmi.ac.in; anirban@jnu.ac.in}

\date{} 

\begin{abstract}
Air pollution poses significant health and environmental challenges, particularly in rapidly urbanizing regions. Delhi-National Capital Region (NCR) experiences severemost air pollution episodes due to complex interactions between anthropogenic emissions and meteorological conditions. Understanding the causal drivers of key pollutants such as fine particulate matter (\( PM_{2.5} \)) and ground-level ozone (\( O_3 \)) is crucial for developing effective mitigation strategies. While previous studies have established correlations between emissions and pollutant levels, they  failed to quantify direct causal effects. This study investigates the causal links of anthropogenic emissions on \( PM_{2.5} \) and \( O_3 \) concentrations using predictive modeling and causal inference techniques. Integrating high-resolution air quality data from January 2018 to August 2023 across 32 monitoring stations, we develop predictive regression models that incorporate meteorological variables (temperature and relative humidity), pollutant concentrations (\( NO_2 \), \( SO_2 \), \( CO \)), and seasonal harmonic components to capture both diurnal and annual cycles.   
 Here, we show that reductions in anthropogenic emissions lead to significant decreases in \( PM_{2.5} \) levels, whereas their effect on \( O_3 \) remains marginal and statistically insignificant. To address spatial heterogeneity, we employ Gaussian Process modeling. Further, we use Granger causality analysis and counterfactual simulation to establish direct causal links. Validation using real-world data from the COVID-19 lockdown confirms that reduced emissions led to a substantial drop in \( PM_{2.5} \) but only a slight, insignificant change in \( O_3 \). 
The findings highlight the necessity of targeted emission reduction policies while emphasizing the need for integrated strategies addressing both particulate and \( O_3 \)  pollution. These insights are crucial for policymakers designing air pollution interventions in worldwide megacities similar to Delhi, and offer a scalable methodology for tackling complex urban air pollution through data-driven decision-making.
\end{abstract}

\begin{document}
\flushbottom
\maketitle

\section{Introduction}

Air pollution remains one of the most pressing environmental and public health challenges for rapidly growing urban centres across the world, and Delhi-NCR stands out as one of the most severely affected regions\cite{deBont2024, Dominici2022, Brewer2023}. The Delhi-NCR experiences frequent and extreme pollution episodes, primarily driven by anthropogenic emissions such as vehicular exhaust, industrial activity, and biomass burning. While many studies have documented trends and seasonal patterns in individual pollutants like \( PM_{2.5} \), \( O_3 \), \( NO_2 \), etc. \cite{ILENIC2024169117,SALONEN2019104887,AKIMOTO1994213,Haiyan2019}, there remains a critical need to understand the causal mechanisms linking emission sources, meteorological factors, and pollutant concentrations in an integrated framework.
Recent research has highlighted the intricate interplay between \( PM_{2.5} \) and ground-level ozone \( O_3 \), noting their complex chemical interactions and joint health impacts. For instance, studies in India and the western United States have found frequent co-occurrence of \( PM_{2.5} \) and \( O_3 \) extremes, amplifying risks to human health \cite{Dmitri2022, urbansci7010009}. Work in New York City and Beijing has demonstrated the difficulties of managing \( O_3 \)  and particulate pollution simultaneously, underscoring the importance of coordinated emission control strategies \cite{zhang2022insights}. In China, a two-pollutant control framework has shown promising results, suggesting that substantial reductions in both \( NO_x \) and Volatile Organic Compounds (VOCs) are needed to effectively mitigate both \( O_3 \) and \( PM_{2.5} \) levels \cite{li2019two}. These findings underscore that targeting a single pollutant is often insufficient and that a systems-level understanding of pollutant interactions is essential\cite{ojha2022mechanisms}.
However, most existing studies are limited by their reliance on correlation-based or univariate approaches, which often fail to account for confounding meteorological factors and feedback mechanisms. There is a growing recognition in the literature that causal inference methods; grounded in counterfactual reasoning and robust statistical frameworks are crucial for establishing the true effects of policy-relevant variables like emission reductions\cite{Geffner_etal_2022}. 

Our aim is to advance this as problem in ``complex system'' by using data science methods\cite{chakrabarti2023data}; a combination of predictive modeling, spatial analysis, and causal inference techniques to quantify the impact of anthropogenic emissions on Delhi's air pollution. This paper develops and applies a set of models that incorporate not only pollutant and meteorological variables but also temporal and spatial structures. We leverage a predictive regression model with harmonic terms to capture both seasonal and diurnal variation, use Gaussian Process modeling to account for spatial correlations, and apply Granger causality to detect lagged directional influences among pollutants. Most importantly, we implement a scenario-based causal analysis; fixing meteorological confounders at typical November levels; to estimate the treatment effect of anthropogenic pollutants (\( NO_2 \), \( SO_2 \), \( CO \)) on both \( PM_{2.5} \) and \( O_3 \). This approach allows us to simulate controlled interventions and evaluate causal effects with greater credibility than correlation-based studies. 
Furthermore, we validate our findings using real-world data from the COVID-19 lockdown in 2020, which acted as a natural experiment by inducing sharp declines in anthropogenic emissions. The comparison of pollution levels during the lockdown (treatment period) with adjacent non-lockdown years (control periods) provides quasi-experimental support for our causal conclusions.
Therefore, our study offers a comprehensive investigation into the dynamics of air pollution in Delhi-NCR. By combining high-resolution data analyses, rigorous statistical modeling, and causal inference techniques, and counterfactual validation, we aim to provide insights that are both scientifically robust and policy-relevant. Our findings not only quantify the role of anthropogenic emissions in shaping air pollution outcomes but also demonstrate the potential benefits— and trade-offs— of emission control strategies targeting multiple pollutants.

\section{Data Description and Exploration}\label{data_explorarion}

For this study, we collected hourly data for Delhi from January 2018 to August 2023. The dataset is publicly accessible through the Central Pollution Control Board (CPCB), Government of India, which can be found at the official  portal: \url{https://cpcb.nic.in/}.  
Air pollutants such as \( NO_2 \), \( SO_2 \), and \( CO \) are major anthropogenic contributors to the levels of \( PM_{2.5} \) and \( O_3 \) \cite{li2019two}. Hence, we study the temporal dynamics of \( PM_{2.5} \) and \( O_3 \) alongside other pollutants (\( NO_2 \), \( SO_2 \), and \( CO \)) as well as meteorological variables (\( AT \) and \( RH \)) over the period from 2018 to 2023.

\begin{figure}[H]
    \centering
    \includegraphics[width=0.22\textwidth]{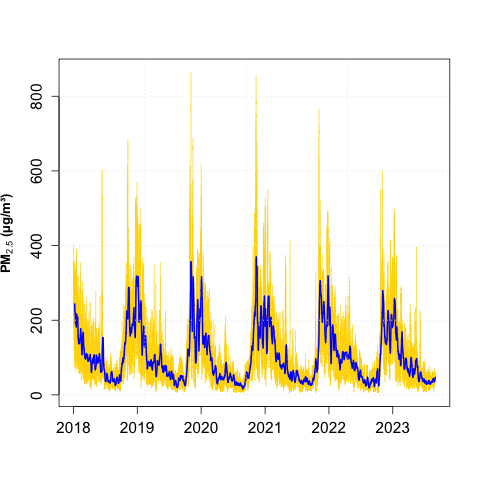}
    \includegraphics[width=0.22\textwidth]{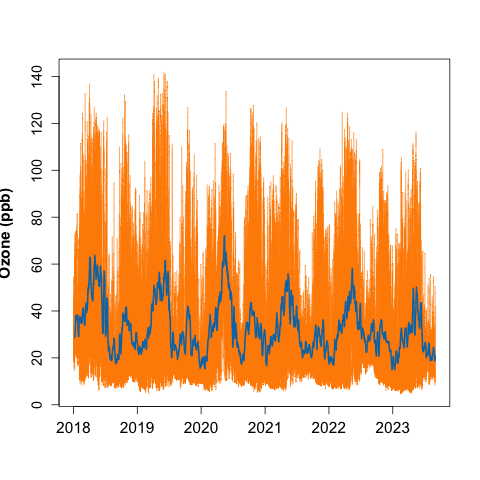}
\includegraphics[width=0.22\textwidth]{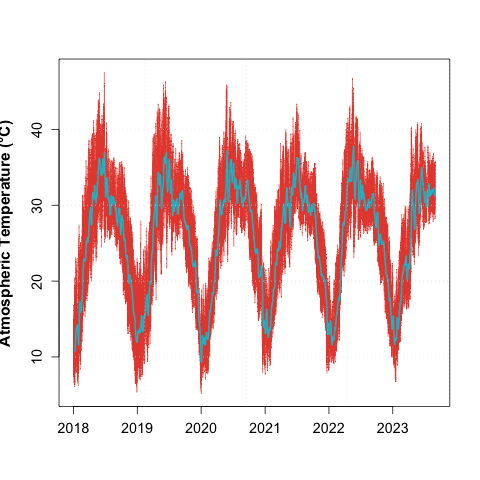}
\includegraphics[width=0.22\textwidth]{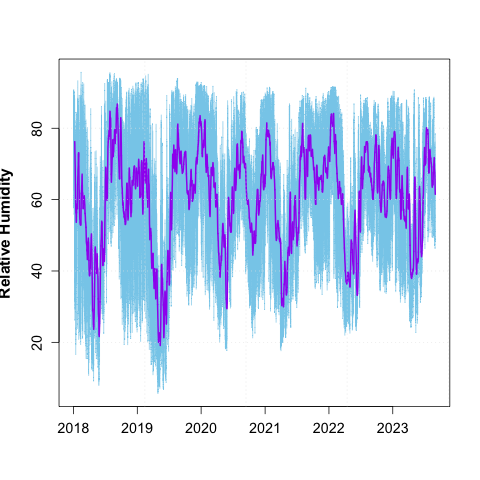}
    \caption{Time series plots of \( PM_{2.5} \), ground-level ozone (\( O_3 \)), atmospheric temperature and relative humidity for Delhi from January 2018 to August 2023. The plots highlight clear annual periodicity in the observed trends.}
    \label{fig:TS_Delhi}
\end{figure}

Figure~\ref{fig:TS_Delhi} presents the time series trends of key air pollutants (\(PM_{2.5}\) and \(O_3\), along with atmospheric temperature and relative humidity in Delhi from January 2018 to August 2023. The upper row highlights the variations in pollutant concentrations, while the lower row captures meteorological influences. A distinct annual periodicity is observed across all variables, indicating the seasonal impact on air pollution. Notably, peaks in \(PM_{2.5}\) correspond to winter months, likely driven by increased emissions and meteorological conditions that trap pollutants closer to the surface. Similarly, \(O_3\) levels exhibit a seasonal cycle, with higher concentrations during warmer months due to photochemical reactions. These trends emphasize the intricate relationship between meteorological factors and air pollution, underscoring the need for season-specific mitigation strategies. 

\begin{figure}[H]
\centering
 {(a)\includegraphics[width=0.22\textwidth]{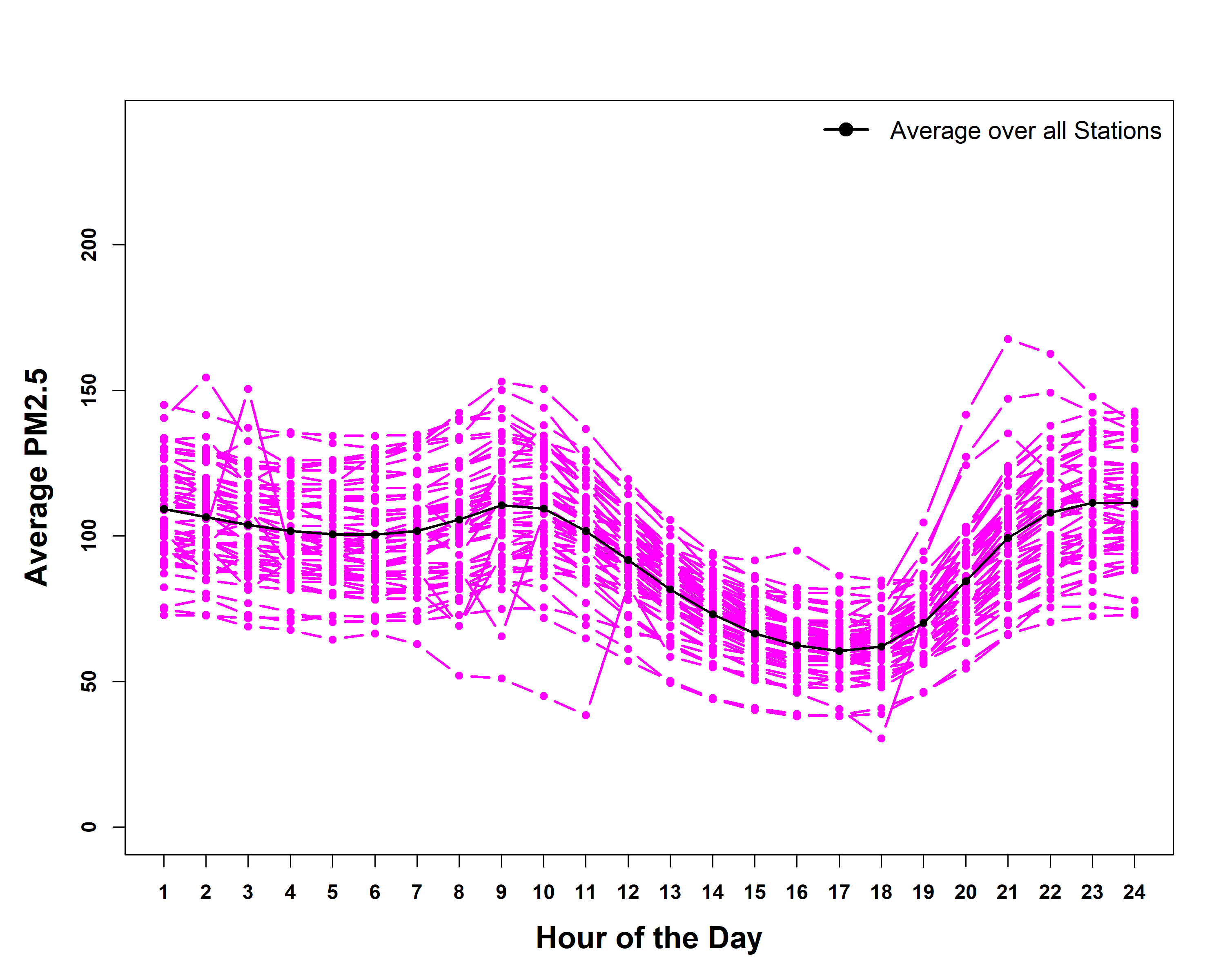}}
 {(b)\includegraphics[width=0.22\textwidth]{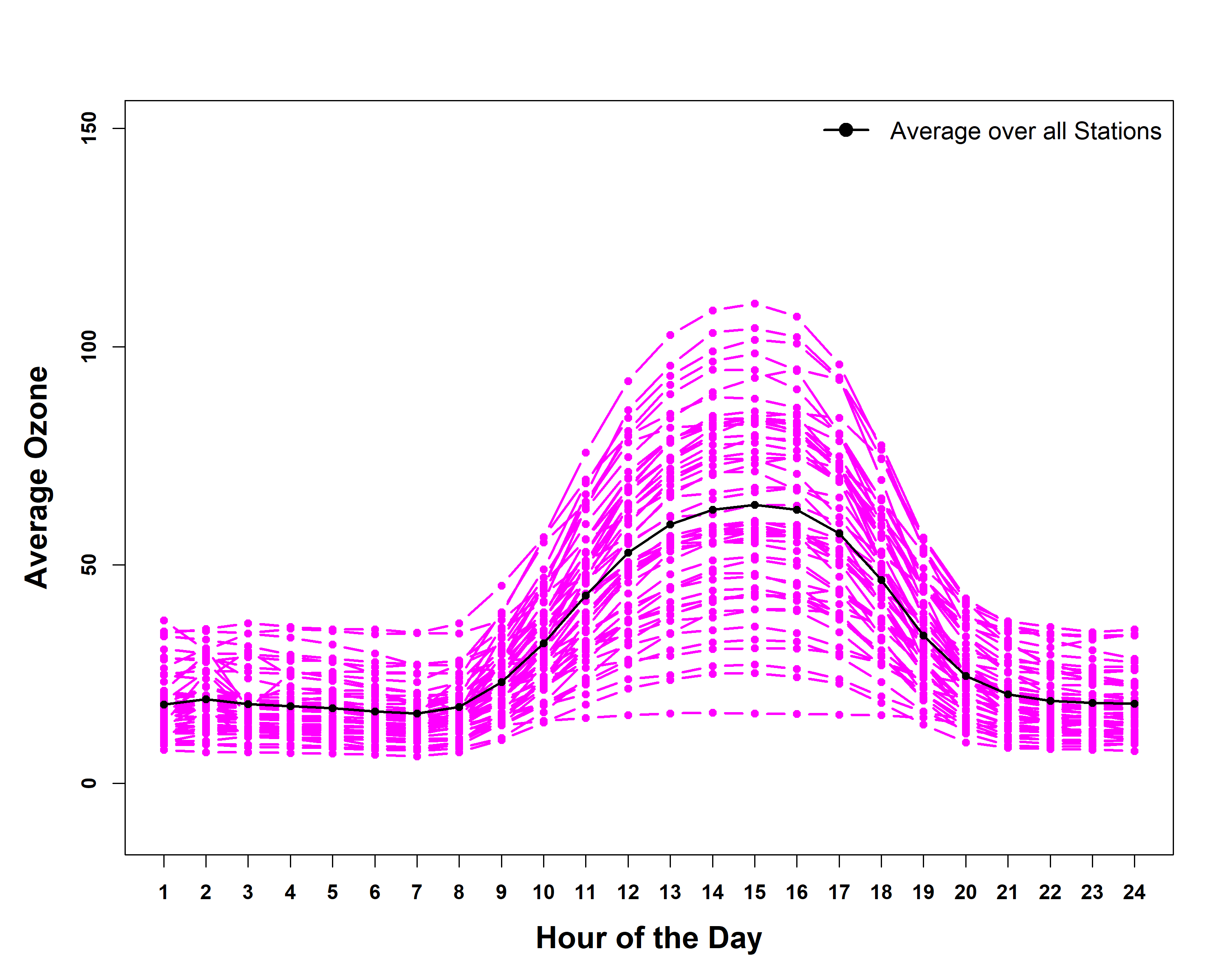}}
 {(c)\includegraphics[width=0.22\textwidth]{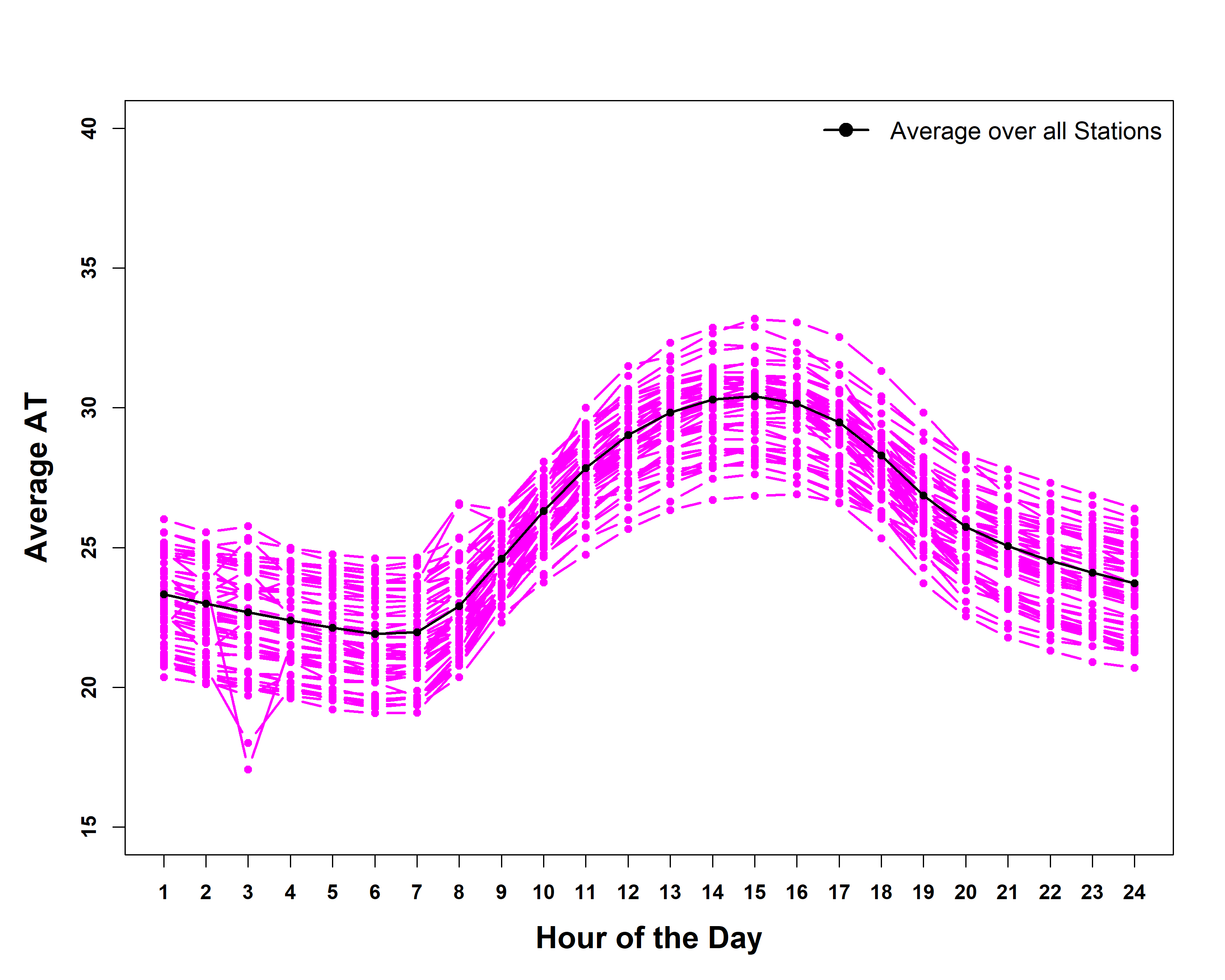}}
 {(d)\includegraphics[width=0.22\textwidth]{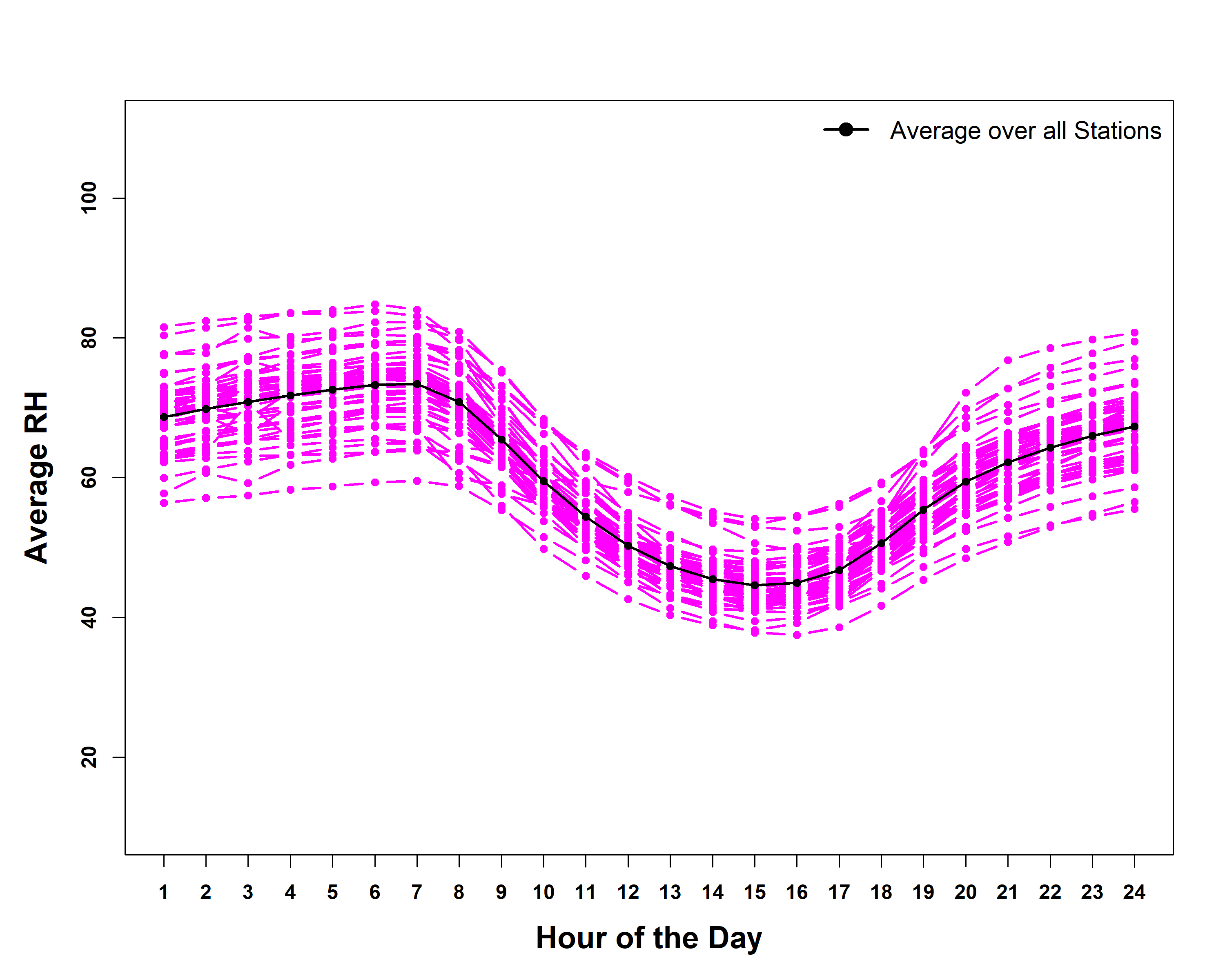}}
\caption{Plots of diurnal variations of the air pollutants and meteorological parameters across 32 locations in Delhi over a
five-year period (January 2018–August 2023): (a) \( PM_{2.5} \) levels are low during the day and high at night;
(b) \( O_3 \) levels peak during the day and drop at night; (c) Temperature is high during the day and low at night; (d) Relative Humidity is low during the day and high at night. The black curve represents the average across all stations. }
\label{fig_delhi_avg_TS}
\end{figure}

\begin{figure}[H]
\centering
 {(a)\includegraphics[width=0.3\textwidth]{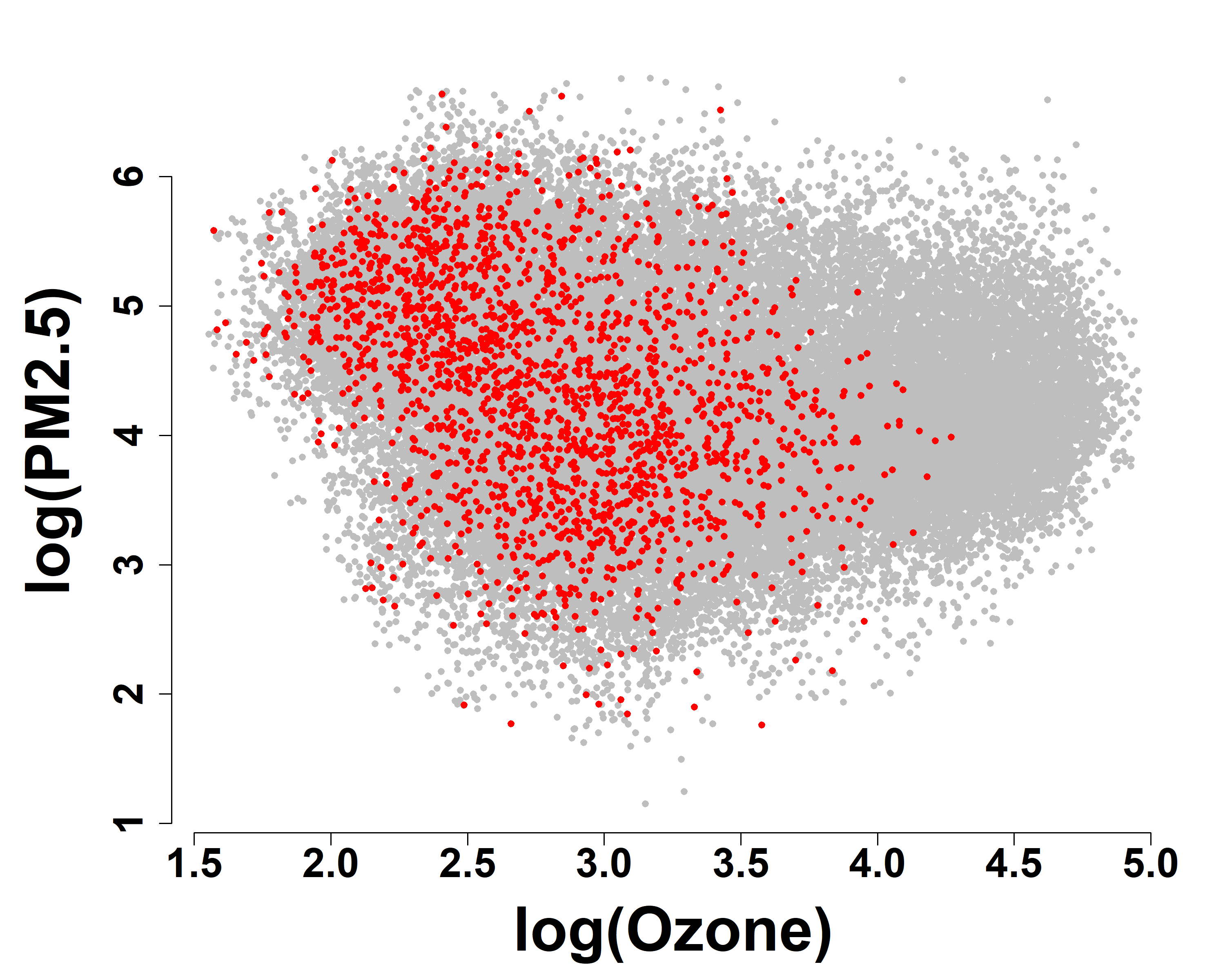}}
 {(b)\includegraphics[width=0.3\textwidth]{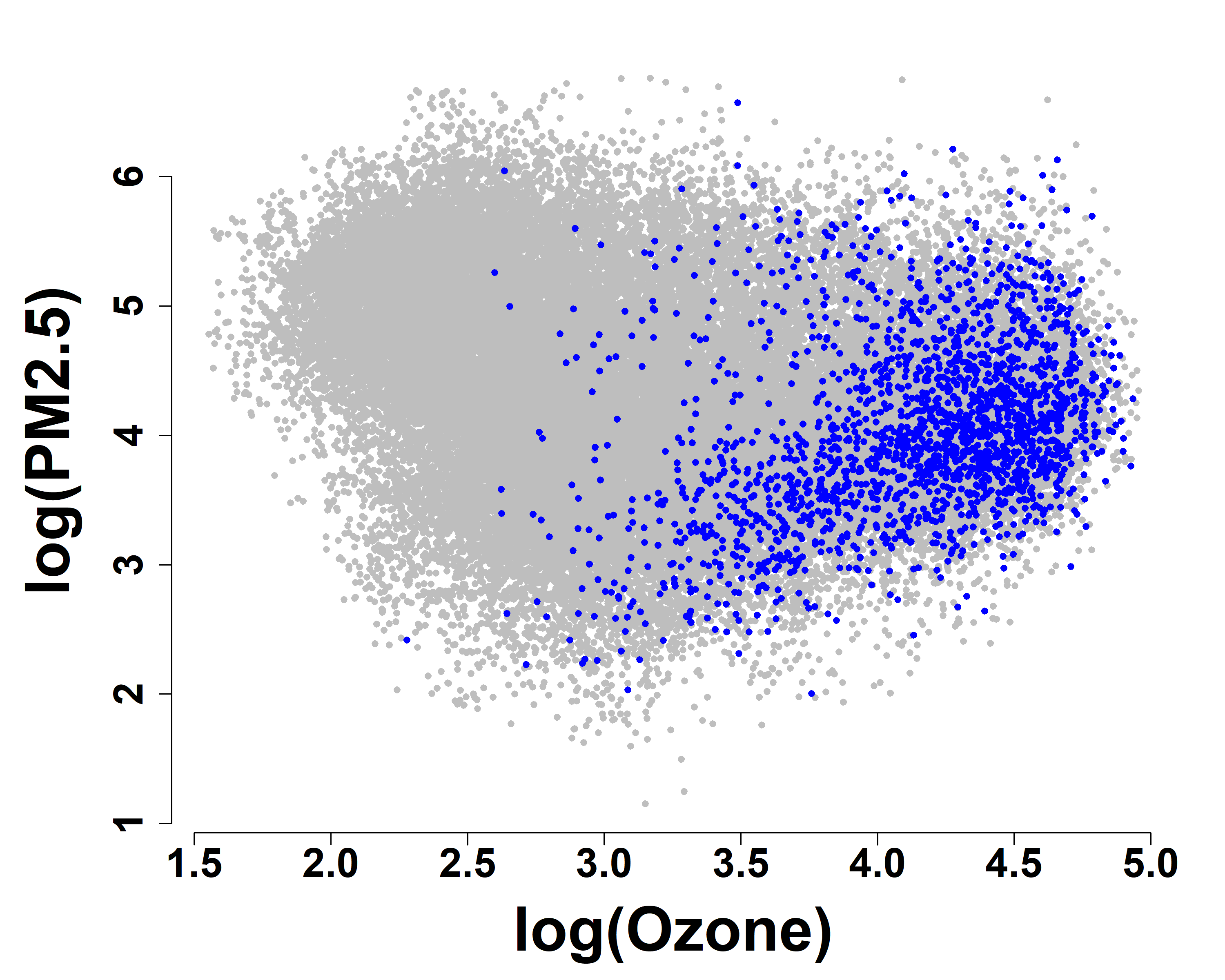}}
 {(c)\includegraphics[width=0.3\textwidth]{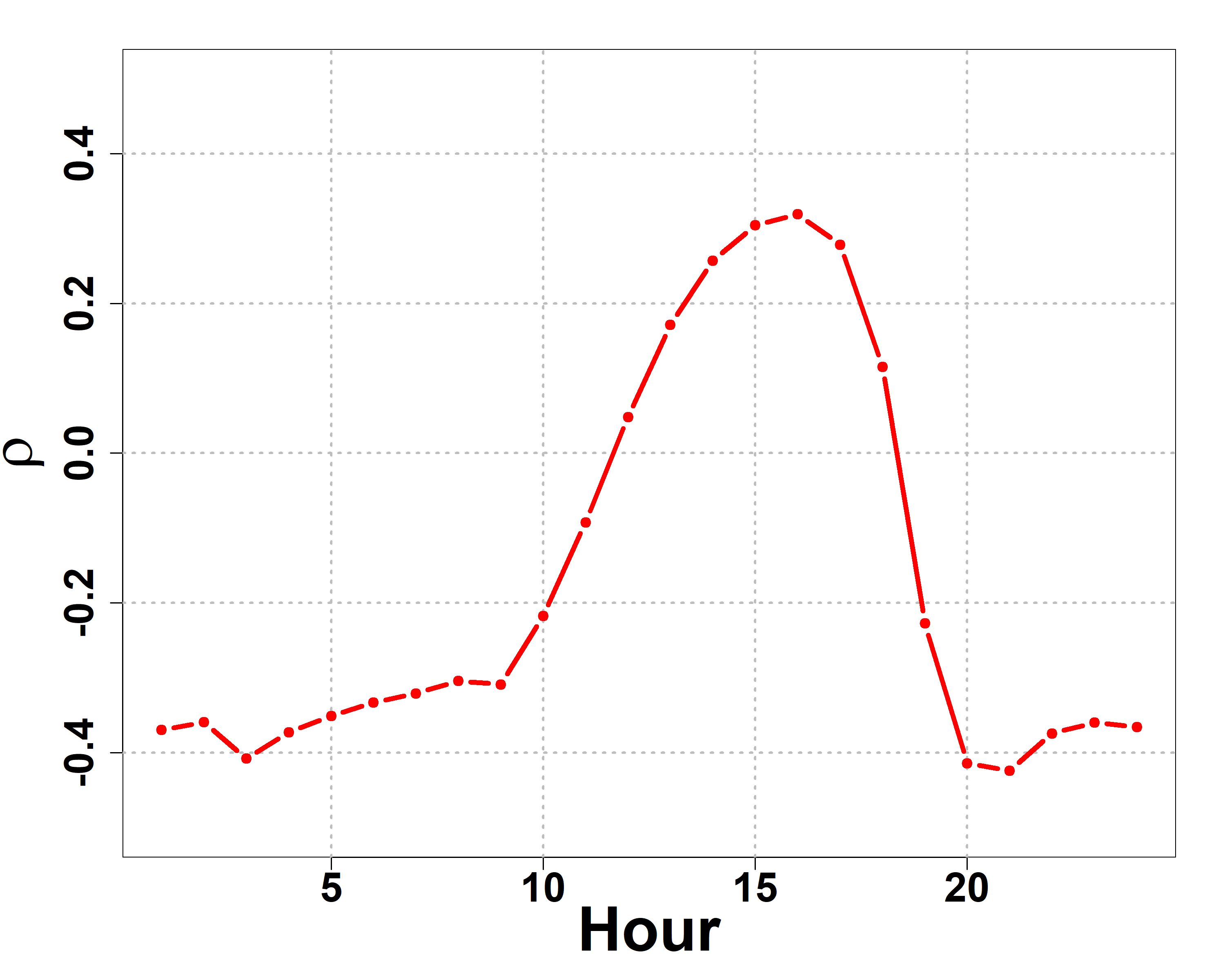}}
\caption{These plots illustrate the correlation between the log-transformed \( PM_{2.5} \) and \( O_3 \)  levels: 
(a) Grey points represent data for all $24$ hours, while red points highlight measurements at $4$ a.m., showing a negative correlation between \( PM_{2.5} \) and \( O_3 \) . 
(b) Blue points indicate data at $4$ p.m., where a clear positive correlation between \( PM_{2.5} \) and \( O_3 \)  is observed. 
(c) This plot depicts the hourly correlation between \( PM_{2.5} \) and \( O_3 \)  across all hours, demonstrating a positive correlation during the day and a negative correlation at night.
}
    \label{fig_delhi_pm2.5_vs_ozone_4am_4pm}
\end{figure}

Figure~\ref{fig_delhi_avg_TS} illustrates the diurnal variations in air pollutants and meteorological parameters across 32 locations in Delhi over a five-year period (January 2018–August 2023). The black curve in each subplot represents the average across all monitoring stations, capturing the overall trend. 
Specifically, Figure~\ref{fig_delhi_avg_TS}(a) shows that $PM_{2.5}$ levels are lower during the day and higher at night, indicating the expansion of the atmospheric boundary layer due to surface heating driven by solar radiation. Figure~\ref{fig_delhi_avg_TS}(b) demonstrates that \( O_3 \)  levels peak during the day and decrease at night due to photochemical production driven by solar radiation. During daylight hours, ozone (\( O_3 \)) is formed through photochemical reactions involving nitrogen oxides (NO\textsubscript{x}) and volatile organic compounds (VOCs) in the presence of sunlight. At night, the absence of solar radiation halts \( O_3 \)  production, and surface deposition, chemical titration by NO, and atmospheric mixing contribute to its decline \cite{li2019two}. Thus, the hourly trends reveal distinct temporal patterns, with \( PM_{2.5} \) concentrations peaking at night, while ground-level ozone (\( O_3 \)) reaches its maximum during the daytime due to photochemical reactions. Atmospheric temperature follows a typical daily cycle, rising during the day and cooling at night, whereas relative humidity exhibits an inverse pattern, increasing at night and decreasing during the daytime. 

Figure~\ref{fig_delhi_pm2.5_vs_ozone_4am_4pm} examines the correlation between log-transformed \( PM_{2.5} \) and \( O_3 \)  levels at different times of the day. The scatter plots indicate a distinct \emph{diurnal pattern} in their relationship: at 4 a.m., a \emph{negative correlation} is observed, suggesting that high \( PM_{2.5} \) levels coincide with low \( O_3 \)  concentrations due to nighttime atmospheric conditions and suppressed photochemical activity. Conversely, at 4 p.m., a \emph{strong positive correlation} emerges, driven by increased solar radiation and photochemical reactions that enhance \( O_3 \)  formation in the presence of fine particulate matter. The overall hourly correlation pattern (Figure~\ref{fig_delhi_pm2.5_vs_ozone_4am_4pm}c) further confirms this trend, demonstrating a \emph{positive correlation during daylight hours and a negative correlation at night}.
These observations highlight the complex interactions between meteorological factors and pollutant levels, particularly the positive correlation between \( O_3 \) and \( PM_{2.5} \) during the day and their negative correlation at night, suggesting the role of secondary chemical processes and boundary layer dynamics in shaping air pollution patterns.

\begin{figure}[H]
\centering
{(a)\includegraphics[width=0.2\textwidth,height=0.2\textwidth]{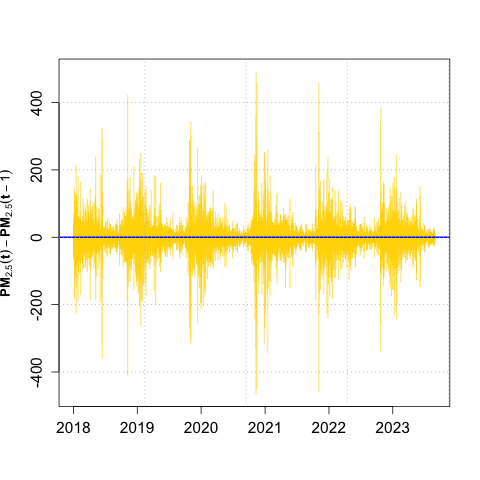}}
{(b)\includegraphics[width=0.2\textwidth,height=0.2\textwidth]{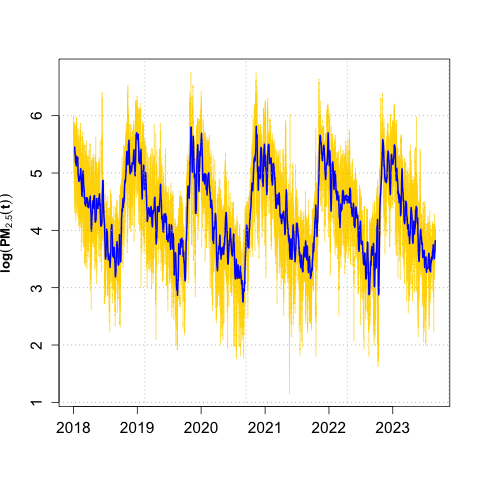}}
{(c)\includegraphics[width=0.2\textwidth,height=0.2\textwidth]{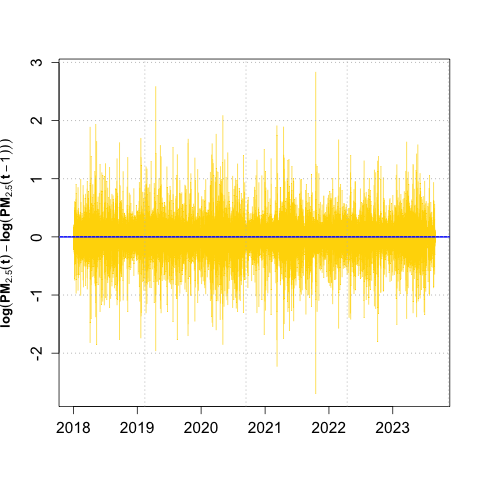}}\\
{(d)\includegraphics[width=0.2\textwidth,height=0.2\textwidth]{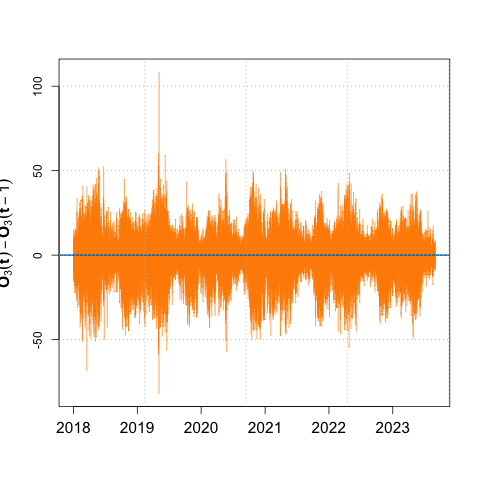}}
{(e)\includegraphics[width=0.2\textwidth,height=0.2\textwidth]{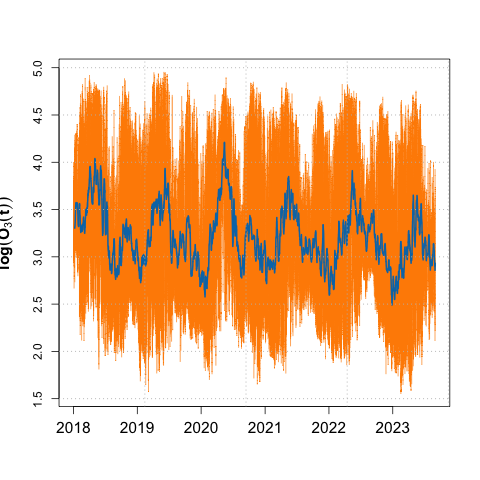}}
{(f)\includegraphics[width=0.2\textwidth,height=0.2\textwidth]{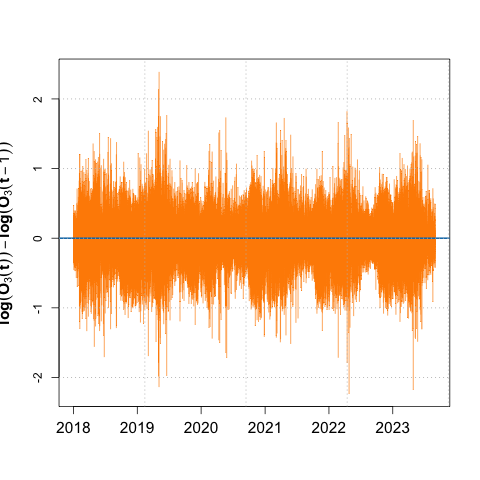}}\\
{(g)\includegraphics[width=0.2\textwidth,height=0.2\textwidth]{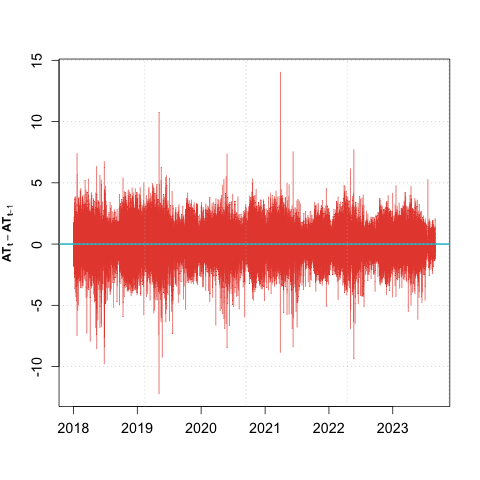}}
{(h)\includegraphics[width=0.2\textwidth,height=0.2\textwidth]{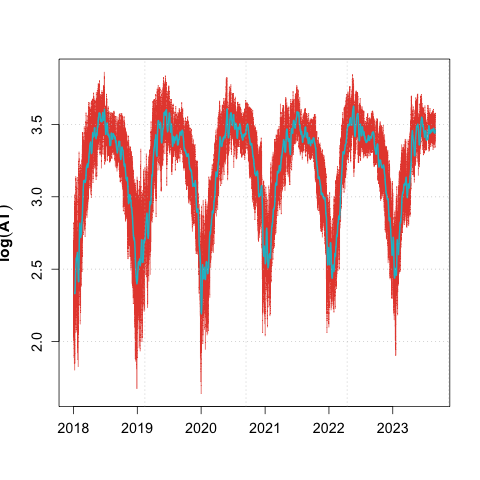}}
{(i)\includegraphics[width=0.2\textwidth,height=0.2\textwidth]{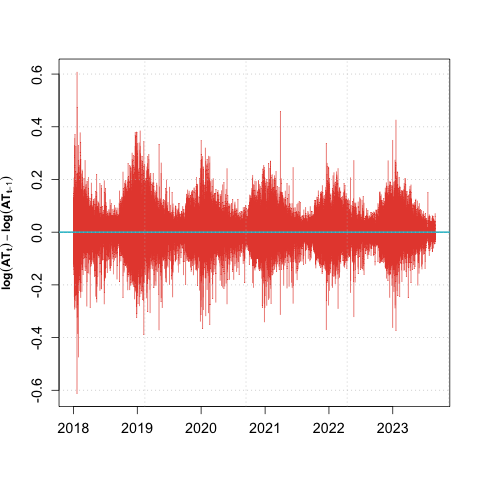}}\\
{(j)\includegraphics[width=0.2\textwidth,height=0.2\textwidth]{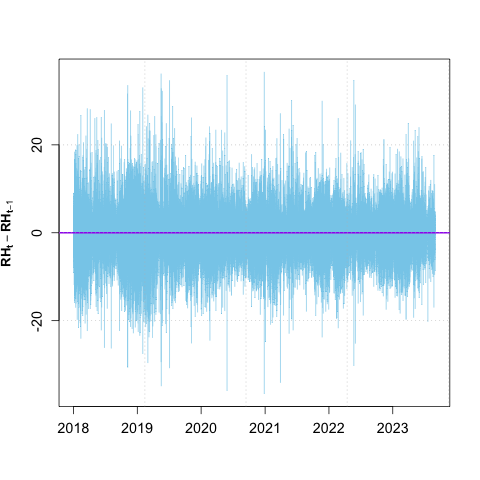}}
{(k)\includegraphics[width=0.2\textwidth,height=0.2\textwidth]{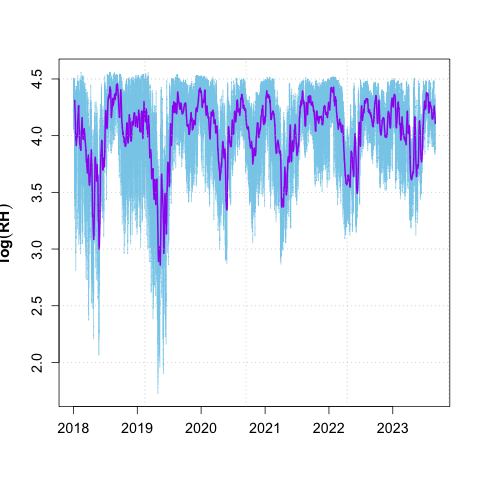}}
{(l)\includegraphics[width=0.2\textwidth,height=0.2\textwidth]{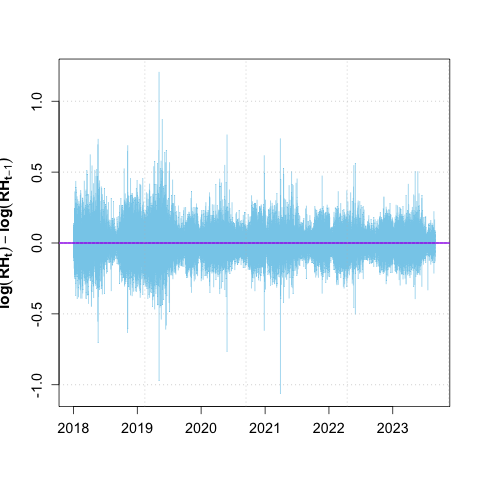}}
\caption{Time series plots illustrating the transformations applied to \( PM_{2.5} \), \( O_3 \), atmospheric temperature, and relative humidity levels in Delhi from January 2018 to August 2023. Each variable is presented in three forms: (a, d, g, j) first differences, capturing short-term fluctuations; (b, e, h, k) logarithmic transformations, stabilizing variance; and (c, f, i, l) first differences of logarithmic transformations, highlighting relative changes over time. These transformations aid in identifying trends, stationarity, and relationships among variables for statistical modeling.}
\label{fig_delhi_TS}
\end{figure}

\begin{figure}[H]
\centering
 {(a)\includegraphics[width=0.2\textwidth,height=0.2\textwidth]{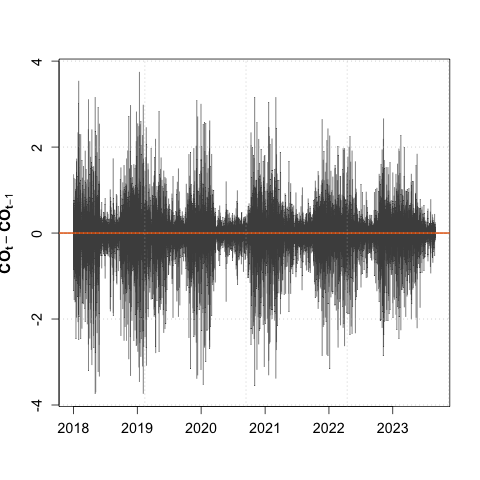}}
 {(b)\includegraphics[width=0.2\textwidth,height=0.2\textwidth]{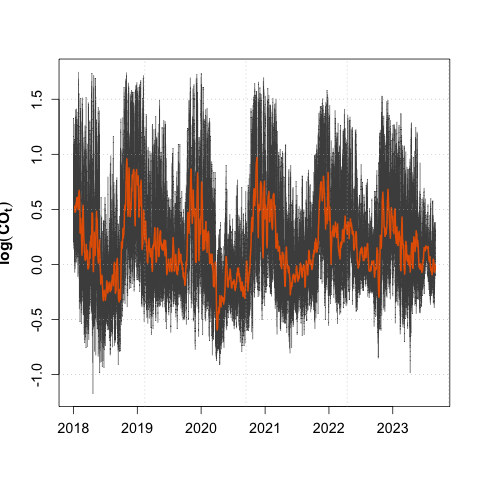}}
 {(c)\includegraphics[width=0.2\textwidth,height=0.2\textwidth]{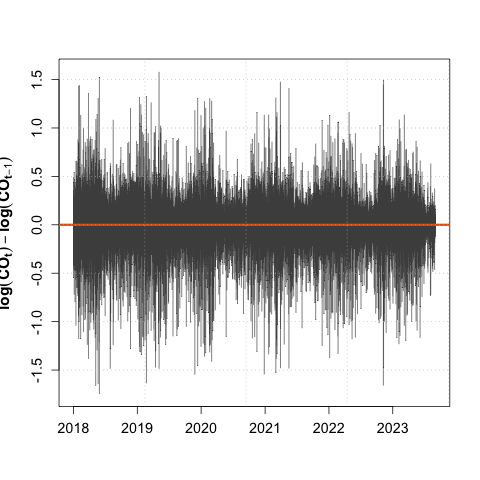}}\\
 {(d)\includegraphics[width=0.2\textwidth,height=0.2\textwidth]{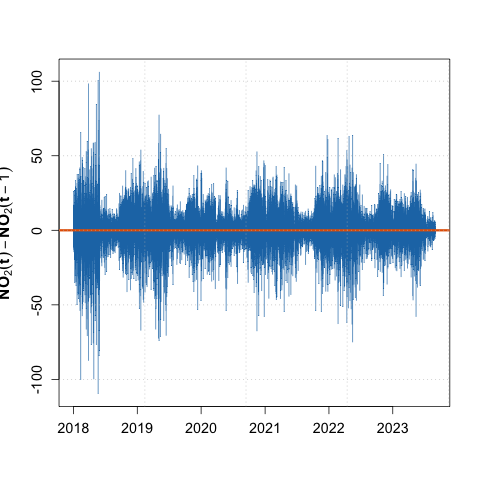}}
 {(e)\includegraphics[width=0.2\textwidth,height=0.2\textwidth]{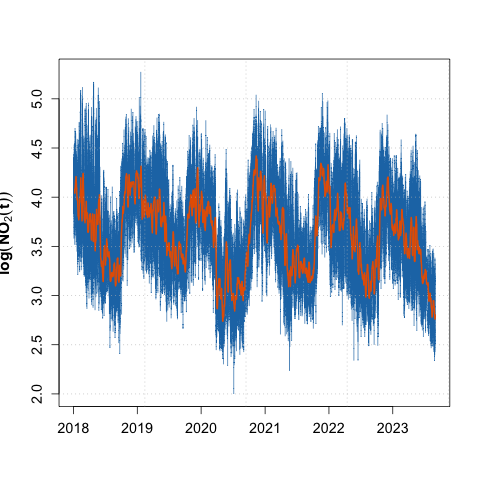}}
 {(f)\includegraphics[width=0.2\textwidth,height=0.2\textwidth]{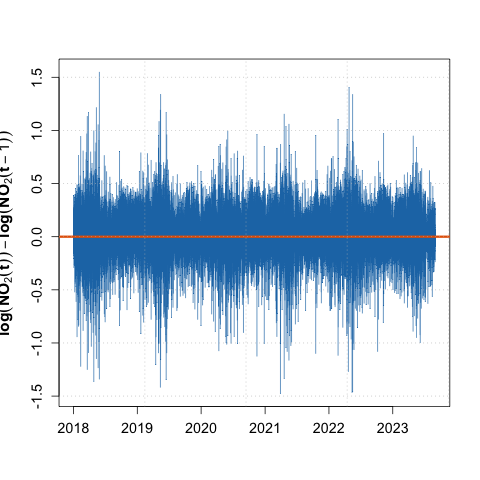}}\\
 {(g)\includegraphics[width=0.2\textwidth,height=0.2\textwidth]{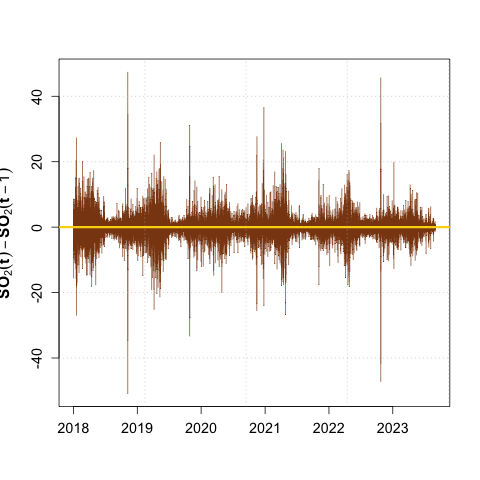}}
 {(h)\includegraphics[width=0.2\textwidth,height=0.2\textwidth]{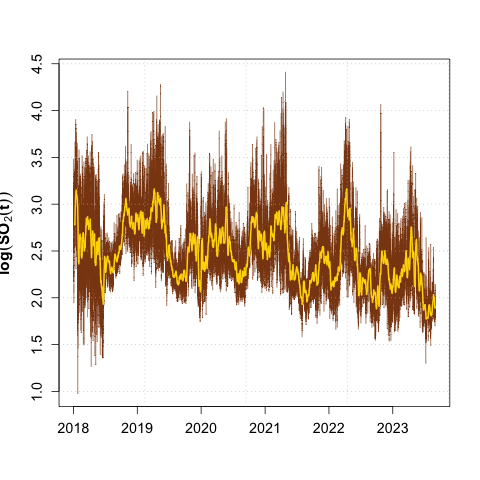}}
 {(i)\includegraphics[width=0.2\textwidth,height=0.2\textwidth]{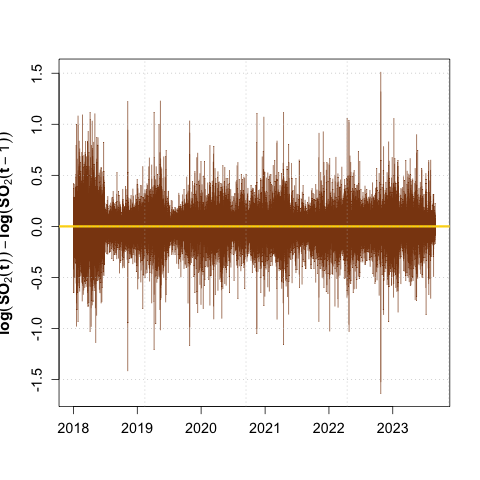}}
\caption{Time series transformations of CO, NO\textsubscript{2}, and SO\textsubscript{2} levels in Delhi from January 2018 to August 2023: (a, d, g) First differences of CO, NO\textsubscript{2}, and SO\textsubscript{2}, capturing short-term fluctuations; (b, e, h) Logarithmic transformations, stabilizing variance and addressing skewness; (c, f, i) First differences of the logarithmic values, highlighting relative changes and ensuring mean stationarity. These transformations aid in identifying trends, stationarity, and relationships among variables for statistical modeling.}
\label{fig_delhi_TS_2}
\end{figure}

Figure \ref{fig_delhi_TS} show the time series plots illustrating the transformations applied to \( PM_{2.5} \), \( O_3 \), atmospheric temperature, and relative humidity levels in Delhi from January 2018 to August 2023. Similarly, 
Figure \ref{fig_delhi_TS_2} show the time series transformations of CO, NO\textsubscript{2}, and SO\textsubscript{2} levels in Delhi from January 2018 to August 2023.
These transformations are applied to stabilize variance, remove trends, and ensure stationarity for statistical modeling. 

\begin{figure}[H]
\centering
 {\includegraphics[width=0.22\textwidth]{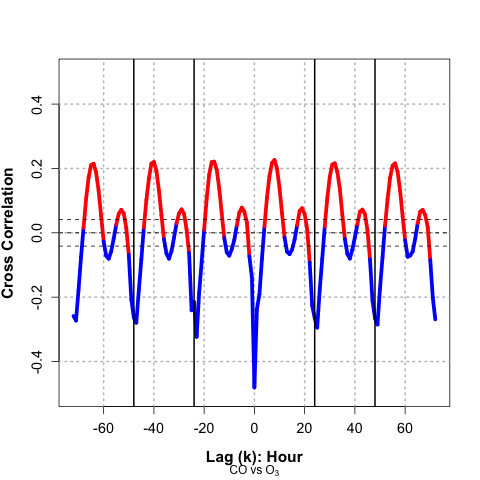}}
 {\includegraphics[width=0.22\textwidth]{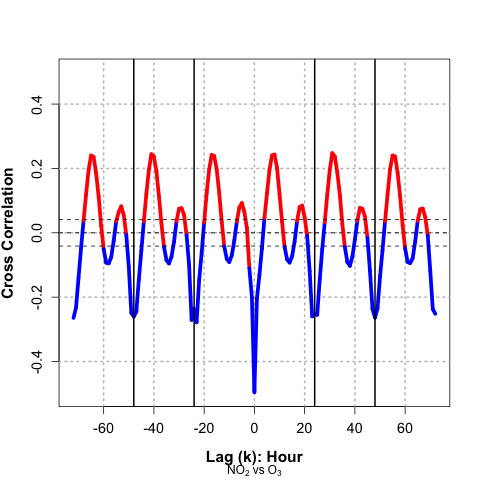}}
 {\includegraphics[width=0.22\textwidth]{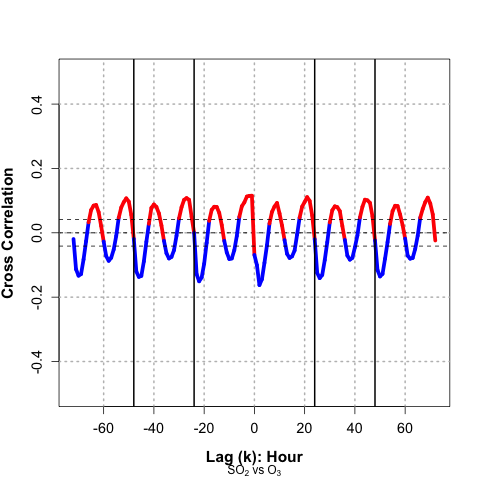}}\\
 {\includegraphics[width=0.22\textwidth]{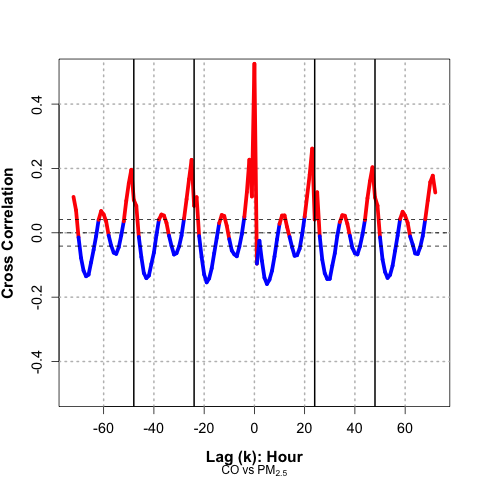}}
 {\includegraphics[width=0.22\textwidth]{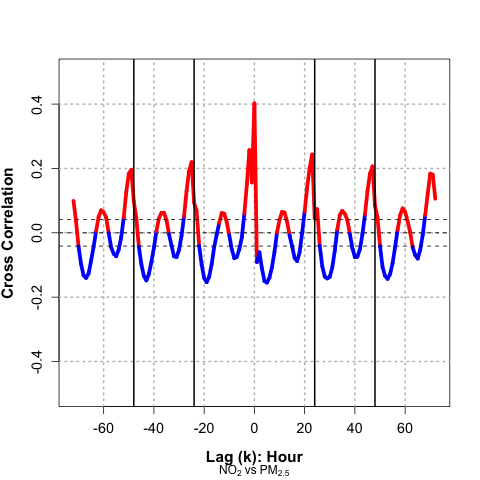}}
 {\includegraphics[width=0.22\textwidth]{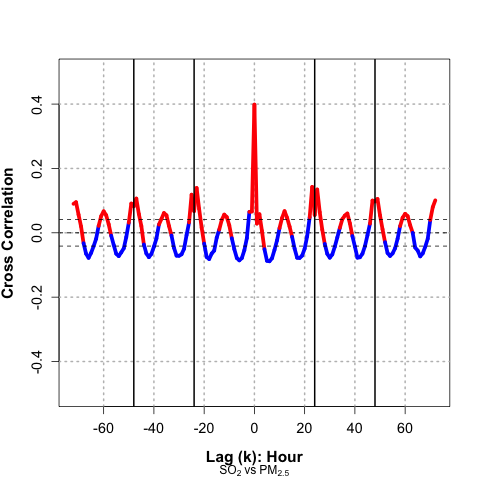}}
\caption{Cross-correlation functions (CCF) between key air pollutants in Delhi from January 2018 to August 2023. The first row illustrates the CCF between CO, NO\textsubscript{2}, and SO\textsubscript{2} with O\textsubscript{3}, while the second row presents the CCF between CO, NO\textsubscript{2}, and SO\textsubscript{2} with \( PM_{2.5} \). These plots help identify the time-lagged relationships between pollutants, revealing how precursor emissions influence secondary pollutant formation over time.}
\label{fig_ccf_delhi}
\end{figure}

Figure \ref{fig_ccf_delhi} presents the cross-correlation functions (CCF) between key air pollutants in Delhi from January 2018 to August 2023, highlighting their time-lagged relationships. The analysis reveals strong diurnal (twice-daily) seasonal patterns in pollutant interactions. \( PM_{2.5} \) shows a strong positive correlation with CO, NO\textsubscript{2}, and SO\textsubscript{2}, indicating their role as primary precursors. In contrast, O\textsubscript{3} exhibits a negative correlation with CO and NO\textsubscript{2}, consistent with photochemical processes where NO\textsubscript{2} plays a crucial role in both O\textsubscript{3} formation and depletion through titration. The impact of SO\textsubscript{2} on O\textsubscript{3} is minimal, suggesting its limited role in direct \( O_3 \)  chemistry. These findings underscore the complex interactions between primary pollutants and secondary formation processes, emphasizing the importance of targeted emission control strategies. These findings reinforce the necessity for data-driven, region-wide pollution control measures that account for both meteorological conditions and emission sources to develop effective mitigation strategies.

\section{Methodology and Results}

The literature suggests that substantial reductions in NO\textsubscript{x} and aromatic VOCs emissions are expected to be highly effective in lowering both \( PM_{2.5} \) and \( O_3 \)  levels \cite{li2019two}. Moreover, studies on the mechanisms and pathways for the coordinated control of \( PM_{2.5} \) and O\textsubscript{3} highlight the complexity of managing these pollutants and underscore the necessity for integrated mitigation strategies \cite{ojha2022mechanisms}. Our exploratory data analysis, presented in Section (\ref{data_explorarion}) indicates that correlations between pollutants, particularly the interplay between \( PM_{2.5} \) and \( O_3 \), emphasize the influence of photochemical reactions and boundary layer dynamics. Hence we focus on development of Statistical models capture this dynamics as function of anthropomorphic pollutants, such as $NO_2$, $CO$ and $SO_2$. We begin by modeling the temporal structure and seasonal variations of air pollution data (see also Ref. \cite{YADAV2024}). 

\subsection{Statistical modeling}

\subsubsection{Predictive Model for Delhi's  $PM_{2.5}$}\label{model:delhi_PM2.5}

The multiple regression model considers \( \log(\text{PM}_{2.5}) \) as the dependent variable, with independent variables comprising the logarithmic transformations of meteorological and air pollutant factors. These include atmospheric temperature (\(\text{AT}\)) and relative humidity (\(\text{RH}\)), along with their squared terms, as well as carbon monoxide (\(\text{CO}\)), nitrogen dioxide (\(\text{NO}_2\)), and sulfur dioxide (\(\text{SO}_2\)), incorporating their squared and interaction terms. Additionally, diurnal (hourly) effects are modeled using sine and cosine functions with a fundamental frequency of \( w_h = \frac{2\pi}{24} \), indexed by \(j\) from 1 to $K$. Similarly, seasonal variations based on the day of the year (\(\text{doy}\)) are represented using sine and cosine terms with a fundamental frequency of \( w_y = \frac{2\pi}{365} \), indexed by \(k\) from 1 to $K$.

The model can be expressed as:

\begin{eqnarray}\label{eqn_modl1}
\log(\text{PM}_{2.5}) &= \beta_0 + \underbrace{\sum_{i=1}^{2} \alpha_i \, \log^i(\text{AT})}_{\text{AT terms}} + \underbrace{\sum_{i=1}^{2} \gamma_i \, \log^i(\text{RH})}_{\text{RH terms}}  + \underbrace{\sum_{i=1}^{2} \delta_i \, \log^i(\text{CO})}_{\text{CO terms}} + \underbrace{\sum_{i=1}^{2} \eta_i \, \log^i(\text{NO}_2)}_{\text{NO}_2 \text{ terms}} + \underbrace{\sum_{i=1}^{2} \theta_i \, \log^i(\text{SO}_2)}_{\text{SO}_2 \text{ terms}} \\
&\quad + \phi_1\, \log(\text{CO})\,\log(\text{NO}_2) + \phi_2\, \log(\text{CO})\,\log(\text{SO}_2) + \phi_3\, \log(\text{NO}_2)\,\log(\text{SO}_2) \nonumber\\
&\quad + \underbrace{\sum_{j=1}^{K} \Bigl[ \lambda_j \, \sin(j \, w_h \, \text{hour}) + \mu_j \, \cos(j \, w_h \, \text{hour}) \Bigr]}_{\text{hourly terms}} + \underbrace{\sum_{k=1}^{K} \Bigl[ \nu_k \, \sin(k \, w_y \, \text{doy}) + \xi_k \, \cos(k \, w_y \, \text{doy}) \Bigr]}_{\text{Seasonal (day-of-year) terms}} \nonumber\\
&\quad + \varepsilon, \nonumber
\end{eqnarray}
where the term \(\beta_0\) denotes the intercept of the model, and the coefficients \(\alpha_i, \gamma_i, \delta_i, \eta_i, \theta_i, \phi_i, \lambda_j, \mu_j, \nu_k,\) and \(\xi_k\) correspond to the various predictors and their interactions, capturing the effects of meteorological variables, pollutant levels, and harmonic components. Finally, \(\varepsilon\) represents the error term, accounting for the unexplained variability in the model. 

\subsubsection{Predictive Model for Delhi's $O_3$}\label{model:delhi_O3}

The meteorological variables in the model include atmospheric temperature (\(\text{AT}\)) and its square, as well as relative humidity (\(\text{RH}\)) and its square. The pollutant variables consist of carbon monoxide (\(\text{CO}\)) and its square, nitrogen dioxide (\(\text{NO}_2\)) and its square, and sulfur dioxide (\(\text{SO}_2\)) and its square. Additionally, the model accounts for interaction effects between these pollutants through the terms \(\log(\text{CO})\log(\text{NO}_2)\), ~~\(\log(\text{CO})\log(\text{SO}_2)\), and \(\log(\text{NO}_2)\log(\text{SO}_2)\) respectively. To capture periodic variations, the model incorporates harmonic terms. Diurnal (hourly) effects are modeled using sine and cosine functions with a fundamental frequency of \( w_h = \frac{2\pi}{24} \), indexed by \(j\) from 1 to $K$. Similarly, seasonal variations based on the day of the year (\(\text{doy}\)) are represented using sine and cosine terms with a fundamental frequency of \( w_y = \frac{2\pi}{365} \), indexed by \(k\) from 1 to $K$.

The model can be expressed as:

\begin{eqnarray}\label{modl_ozone}
\log(\text{Ozone}) &= \beta_0 + \underbrace{\sum_{i=1}^{2} \alpha_i \, \log^i(\text{AT})}_{\text{AT terms}} + \underbrace{\sum_{i=1}^{2} \gamma_i \, \log^i(\text{RH})}_{\text{RH terms}}  + \underbrace{\sum_{i=1}^{2} \delta_i \, \log^i(\text{CO})}_{\text{CO terms}} + \underbrace{\sum_{i=1}^{2} \eta_i \, \log^i(\text{NO}_2)}_{\text{NO}_2 \text{ terms}} + \underbrace{\sum_{i=1}^{2} \theta_i \, \log^i(\text{SO}_2)}_{\text{SO}_2 \text{ terms}} \\
&\quad + \phi_1\, \log(\text{CO})\,\log(\text{NO}_2) + \phi_2\, \log(\text{CO})\,\log(\text{SO}_2) + \phi_3\, \log(\text{NO}_2)\,\log(\text{SO}_2) \nonumber\\
&\quad + \underbrace{\sum_{j=1}^{K} \Bigl[ \lambda_j \, \sin(j \, w_h \, \text{hour}) + \mu_j \, \cos(j \, w_h \, \text{hour}) \Bigr]}_{\text{hourly terms}} + \underbrace{\sum_{k=1}^{K} \Bigl[ \nu_k \, \sin(k \, w_y \, \text{doy}) + \xi_k \, \cos(k \, w_y \, \text{doy}) \Bigr]}_{\text{Seasonal (day-of-year) terms}} \nonumber\\
&\quad + \varepsilon, \nonumber
\end{eqnarray}
where the term \(\beta_0\) denotes the intercept of the model, and the coefficients \(\alpha_i, \gamma_i, \delta_i, \eta_i, \theta_i, \phi_i, \lambda_j, \mu_j, \nu_k,\) and \(\xi_k\) correspond to the various predictors and their interactions, capturing the effects of meteorological variables, pollutant levels, and harmonic components. Finally, \(\varepsilon\) represents the error term, accounting for the unexplained variability in the model. 

\subsubsection{Ridge Regression}
These models account for the influence of meteorological factors, pollutant concentrations, and periodic patterns (both diurnal and seasonal) on \(\text{PM}_{2.5}\) and \( O_3 \) levels, respectively. 
To address multicollinearity among the predictor variables, we applied Ridge regression as a regularisation technique. The optimal penalty parameter (\( \lambda \)) was selected using 10-fold cross-validation. This Ridge-corrected model helps stabilise coefficient estimates and improves generalisability without compromising interpretability. All subsequent scenario analyses and causal inference were carried out using the coefficients derived from this regularised model.

\begin{figure}[H]
\centering
{\includegraphics[width=0.45\textwidth,height=0.35\textwidth]{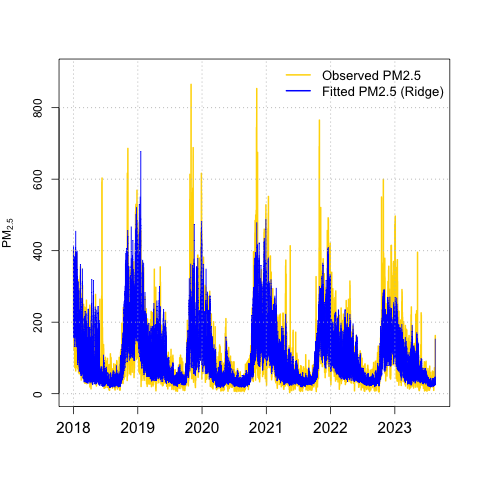}} {\includegraphics[width=0.45\textwidth,height=0.35\textwidth]{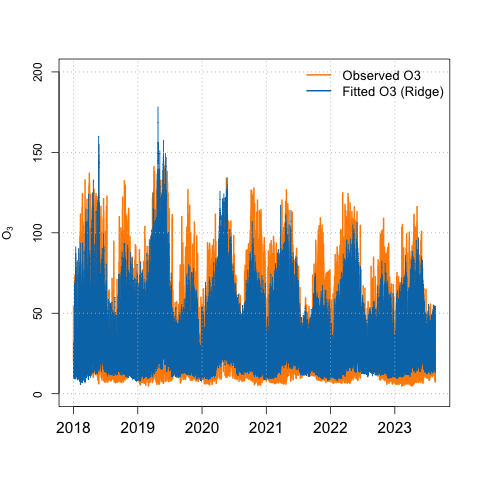}}
\caption{Plots of observed versus predicted values for \( PM_{2.5} \) and \( O_3 \)  from Models \ref{eqn_modl1} and \ref{modl_ozone}  at Delhi. The R-squared value is 0.8194 for \( PM_{2.5} \) and 0.7585 for \( O_3 \) .}
    \label{fig_delhi_model_result}
\end{figure}

Using out-of-sample RMSE values, optimizing the periodic order ($K$) yields: (a) For \( PM_{2.5} \),  $K=8$. (b) For \( O_3 \), $K=5$. The RMSE remains relatively constant beyond these values. 
Figure \ref{fig_delhi_model_result} presents the observed versus predicted values for \( PM_{2.5} \) and \( O_3 \)  based on Model \ref{eqn_modl1} and Model \ref{modl_ozone}, with R-squared values of 81.94\% for \( PM_{2.5} \) and 75.85\% for \( O_3 \) .

\begin{figure}[H]
    \centering
    \begin{tabular}{ccc}
        \includegraphics[width=0.32\linewidth,height=0.22\linewidth]{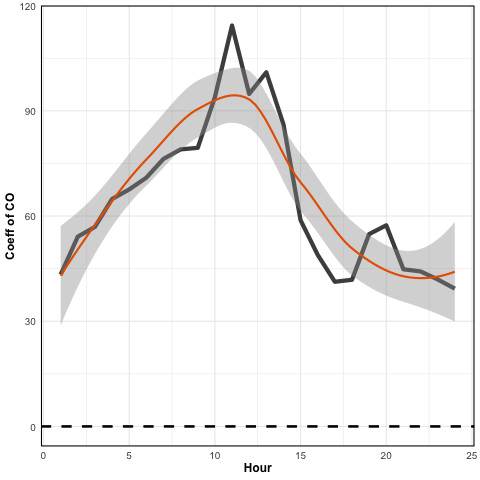}  
       & 
       \includegraphics[width=0.32\linewidth,height=0.22\linewidth]{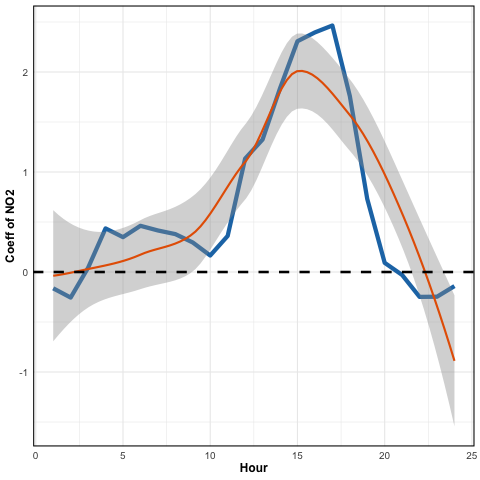} 
       &
       \includegraphics[width=0.32\linewidth,height=0.22\linewidth]{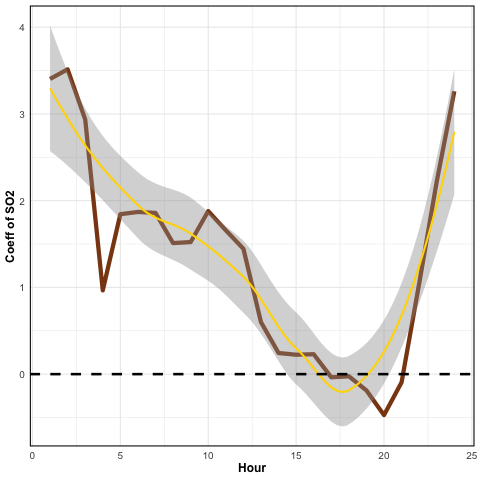}        
       \\
        \includegraphics[width=0.32\linewidth,height=0.22\linewidth]{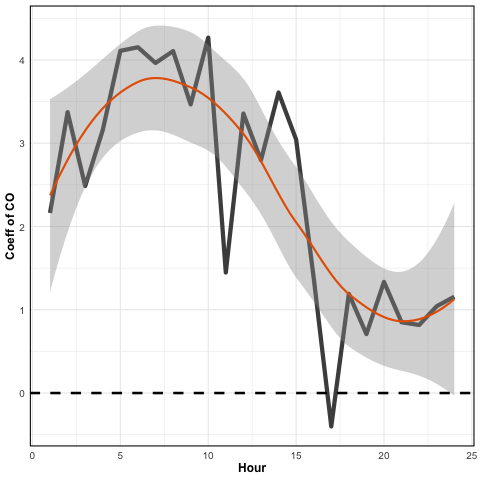}  
       & 
       \includegraphics[width=0.32\linewidth,height=0.22\linewidth]{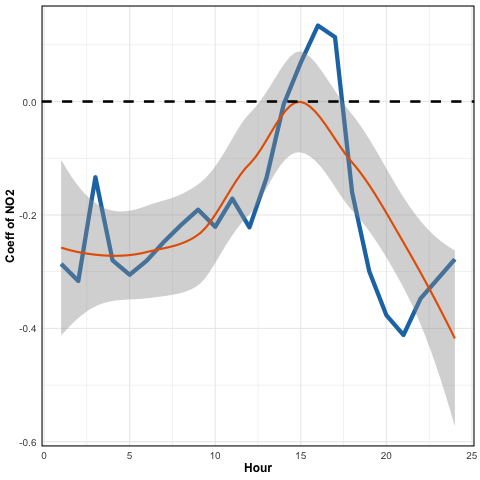} 
       & 
       \includegraphics[width=0.32\linewidth,height=0.22\linewidth]{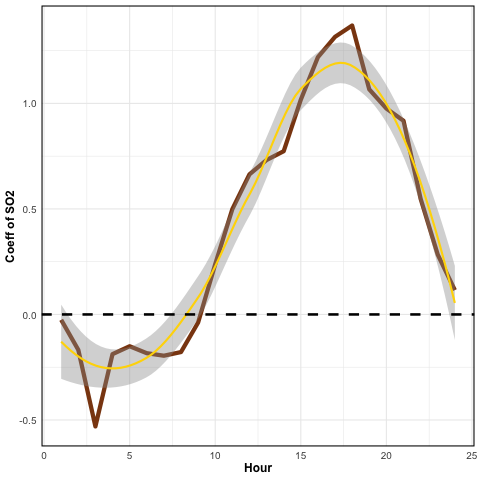}    \\
    \end{tabular}
    \caption{Hourly variation in the estimated coefficients of CO, NO\textsubscript{2}, and SO\textsubscript{2} on \( PM_{2.5} \) (top row) and O\textsubscript{3} (bottom row) in Delhi. The analysis is based on hourly regression models fitted using data from January 2018 to August 2023. The top row shows how the contribution of each pollutant to \( PM_{2.5} \) concentrations varies by hour, while the bottom row illustrates their influence on ground-level \( O_3 \) . These patterns reflect the underlying photochemical and atmospheric processes that govern pollutant formation and dispersion over the course of the day.}
    \label{fig:hourly_coefficients}
\end{figure}

The dynamic variation of the coefficients is illustrated in Figure \ref{fig:hourly_coefficients}.
The top row shows how the contribution of each pollutant to $PM_{2.5}$ concentrations varies by hour, while the bottom row illustrates their influence on ground-level \( O_3 \) .
Hourly coefficient plots provided additional nuance, showing that the strength and direction of each pollutant’s effect on \( PM_{2.5} \) and \( O_3 \) vary throughout the day. 
These patterns reflect the underlying photochemical and atmospheric processes that govern the formation and dispersion of pollutants throughout the day. $NO_2$ generally shows a positive effect on $PM_{2.5}$ throughout the day, while its effect on $O_3$ is negative at night, consistent with the NO$_x$ titration mechanism. $SO_2$ tends to increase $PM_{2.5}$ levels during nighttime and has a positive influence on $O_3$ during daytime hours. $CO$ exhibits a positive effect on both $PM_{2.5}$ and $O_3$, reflecting its role as a precursor in secondary pollutant formation.

\subsection{Granger Causal Inference and Correction for spatial correlations using Gaussian Process}

\subsubsection{Granger causality test}

The Granger causality test \cite{Granger1969} determines whether one time series can predict another by assessing whether past values of one variable provide statistically significant information about the future values of another. For example, we investigate whether \( NO_2 \) can predict \( PM_{2.5} \) by assessing whether past values of \( NO_2 \) provide statistically significant information about the future values of \( PM_{2.5} \). While the test does not establish true causation, it identifies predictive relationships based on lagged correlations in time-series data. A fundamental requirement for applying the test is that the time series must be \emph{stationary}, meaning their statistical properties (mean, variance, and autocorrelation) should remain constant over time. If the series are non-stationary, transformations such as differencing or cointegration techniques must be applied to ensure valid results. In Section \ref{data_explorarion}, Figure \ref{fig:TS_Delhi} shows that air pollutant levels over the years are non-stationary, as air pollution levels are significantly higher during the winter season compared to the monsoon season. However, Figure \ref{fig_delhi_TS} demonstrates that the log differences of the time series are stationary. To construct the Granger causality test, we use the vector autoregressive (VAR) model based on \emph{log differences} of the time series, formulated as follows:
\[
\Delta \log(Y_t) = \alpha + \sum_{i=1}^{p} \beta_i \Delta \log(Y_{t-i}) + \sum_{j=1}^{p} \gamma_j \Delta \log(X_{t-j}) + \varepsilon_t,
\]
where \( \Delta \log(Y_t) = \log(Y_t) - \log(Y_{t-1}) \) and \( \Delta \log(X_t) = \log(X_t) - \log(X_{t-1}) \). The Granger causality hypothesis test is then formulated as follows:
\begin{itemize}
    \item \emph{Null Hypothesis} (\( H_0 \)): \( \Delta \log(X_t) \) does not \emph{Granger-cause} \( \Delta \log(Y_t) \), i.e.,
    \[
    \gamma_1 = \gamma_2 = \dots = \gamma_p = 0,
    \]
    meaning all lagged coefficients of \( \Delta \log(X_t) \) are zero.

    \item \emph{Alternative Hypothesis} (\( H_A \)): \( \Delta \log(X_t) \) \emph{Granger-causes} \( \Delta \log(Y_t) \), i.e., at least one
    \[
    \gamma_j \neq 0, \quad \text{for some } j \in \{1,2,\dots,p\}.
    \]
\end{itemize}
To test this hypothesis, we conduct an \emph{F-test} to determine whether the inclusion of lagged values of \( X_t \) significantly improves the prediction of \( Y_t \).

Table~\ref{tab:granger_causal} presents the $F$-values from Granger causality tests assessing whether concentrations of key anthropogenic pollutants; \( NO_2 \), \( CO \), and \( SO_2 \); can statistically predict future values of two major air pollution indicators: \( PM_{2.5} \) and \( O_3 \), over five consecutive years (2018–2022). The analysis is separated by season: monsoon (July - September) and winter (November - January), which exhibit sharply contrasting meteorological and pollution dynamics. Across all years, the results show a clear seasonal effect: the causal influence of pollutants on \( PM_{2.5} \) and \( O_3 \) is consistently stronger in winter compared to the monsoon. For instance, in winter 2022, the $F$-values for \( NO_2 \rightarrow PM_{2.5} \) and \( CO \rightarrow PM_{2.5} \) reached 89.47 and 65.23 respectively—both substantially higher than their monsoon counterparts. This pattern is consistent with known winter meteorological phenomena in Delhi, including lower atmospheric boundary layers, stagnant air, and increased residential and industrial emissions, all of which enhance the persistence and accumulation of fine particulate matter.

During the monsoon season, the $F$-values for most pollutant–outcome relationships are noticeably weaker. Rainfall, higher humidity, and enhanced boundary layer mixing likely reduce pollutant concentrations and dilute their predictive relationships with \( PM_{2.5} \) and \( O_3 \). For example, in the monsoon of 2022, \( SO_2 \rightarrow PM_{2.5} \) had an $F$-value of only 0.69, suggesting minimal or no predictive power. However, there is substantial inter-annual variation, even within seasons. In the winter of 2020, \( NO_2 \rightarrow PM_{2.5} \) had an $F$-value of 76.81, while in 2019, it dropped to 31.14; potentially reflecting differences in pollution control measures, meteorological anomalies, or post-COVID emission patterns. Notably, \( CO \) consistently shows strong predictive power for \( PM_{2.5} \) during winter, with peak $F$-values in 2020 and 2021, aligning with its origin from incomplete combustion and vehicular emissions, which intensify under winter heating demands and reduced dispersion. For \( O_3 \), the influence of \( NO_2 \) and \( SO_2 \) appears more complex and season-dependent. While \( NO_2 \) shows moderate to high $F$-values for \( O_3 \) in winter; reflecting its role in both \( O_3 \)  production and depletion through titrationl; the effect of \( SO_2 \) is significant in some years (e.g., 2018 and 2022) but negligible in others, suggesting episodic industrial or regional influences. Overall, this table provides strong evidence that the Granger causal influence of air pollutants on Delhi's air pollution varies substantially across seasons and years, pointing to the importance of time-sensitive and pollutant-specific regulatory strategies. These insights also reinforce the necessity for integrating meteorological context and long-term monitoring in causal modeling frameworks for urban air pollution.

\begin{table}[H]
\centering
\renewcommand{\arraystretch}{1.2} 
\setlength{\tabcolsep}{8pt} 
\begin{tabular}{lccccc}
  \hline
  \textbf{Relations}  & \textbf{2018} & \textbf{2019} & \textbf{2020} & \textbf{2021} & \textbf{2022} \\ 
  \hline
  \multicolumn{6}{c}{\textbf{July to September (Monsoon)}} \\
  \hline
  $NO_2 \rightarrow PM_{2.5}$  & 39.13 & 34.00 & 39.70 & 36.54 & 30.62 \\ 
  $CO \rightarrow PM_{2.5}$  & 43.29 & 30.52 & 28.89 & 47.10 & 31.10 \\ 
  $SO_2 \rightarrow PM_{2.5}$  & 3.33 & 3.84 & 5.76 & 13.62 & 0.69 \\ 
  $NO_2 \rightarrow O_3$  & 6.26 & 9.49 & 3.43 & 19.51 & 13.57 \\ 
  $CO \rightarrow O_3$  & 2.90 & 2.86 & 8.80 & 10.98 & 5.75 \\ 
  $SO_2 \rightarrow O_3$  & 2.01 & 5.47 & 6.10 & 4.48 & 1.82 \\ 
  \hline
  \multicolumn{6}{c}{\textbf{November to January (Winter)}} \\
  \hline
  $NO_2 \rightarrow PM_{2.5}$  & 65.04 & 31.14 & 76.81 & 74.34 & 89.47 \\ 
  $CO \rightarrow PM_{2.5}$  & 96.68 & 31.34 & 84.42 & 108.80 & 65.23 \\ 
  $SO_2 \rightarrow PM_{2.5}$  & 35.37 & 12.13 & 10.11 & 13.35 & 20.12 \\ 
  $NO_2 \rightarrow O_3$ & 43.66 & 17.12 & 16.68 & 36.03 & 16.31 \\ 
  $CO \rightarrow O_3$ & 2.42 & 5.89 & 4.00 & 5.37 & 7.81 \\ 
  $SO_2 \rightarrow O_3$  & 44.32 & 39.99 & 42.73 & 15.30 & 39.79 \\ 
  \hline
\end{tabular}
\caption{$F$-Values of Granger causality test}
\label{tab:granger_causal}
\end{table}

\subsubsection{Modelling Spatial Correlation using Gaussian Process} \label{sec:model_spatial}

\begin{figure}[H]
	\centering
	\includegraphics[width=0.4\textwidth]{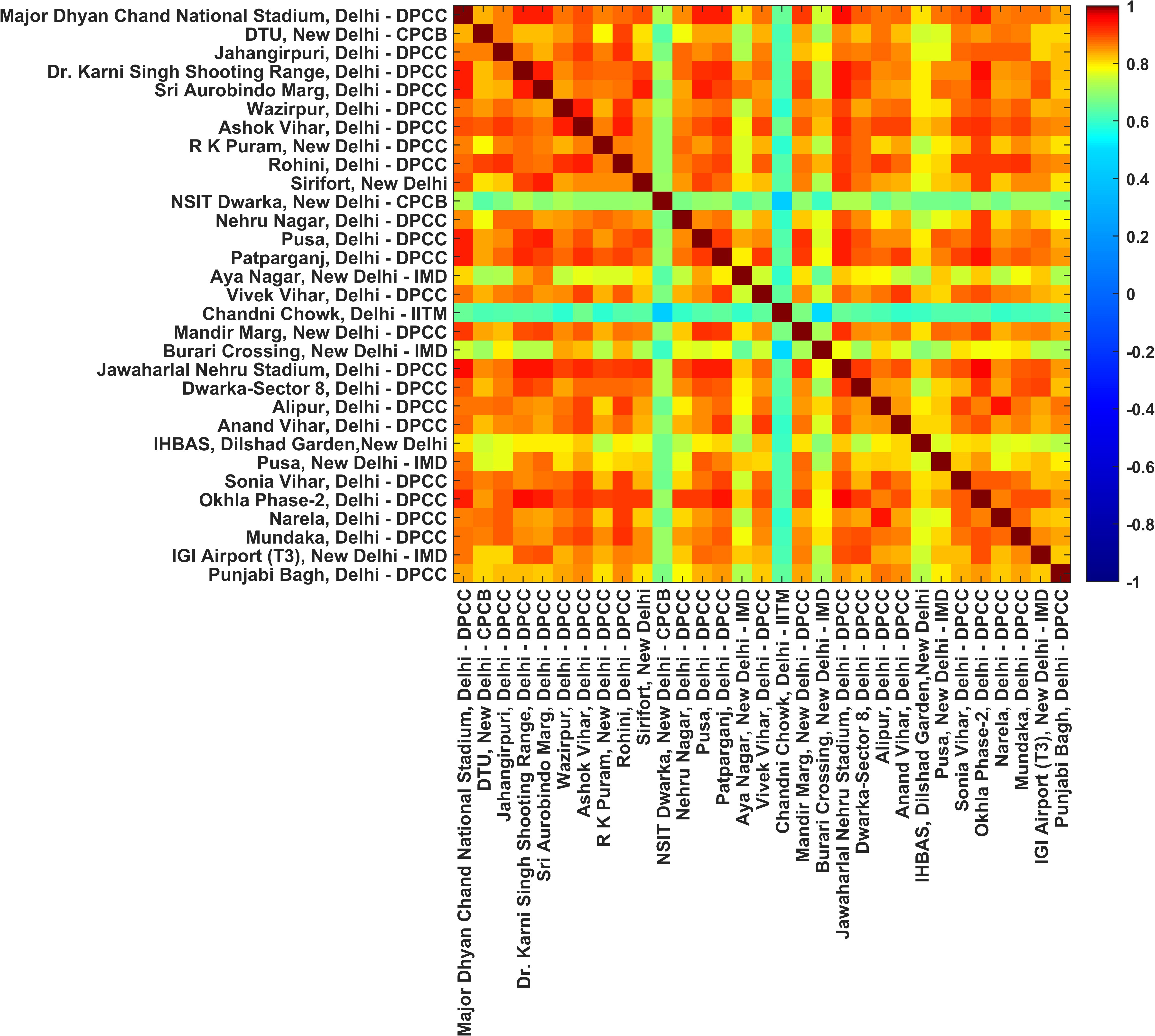}
	\caption{Spatial Correlation plot of \( PM_{2.5} \) levels across Delhi's 32 locations reveals that these locations are highly correlated with one another. }
	\label{delhi_correlation_matrix}
\end{figure}

Figure~\ref{delhi_correlation_matrix} presents the spatial correlation matrix of \( PM_{2.5} \) levels across 32 locations in Delhi. The high correlation values indicate that air pollution levels are strongly interconnected across different monitoring sites, suggesting that pollution is influenced by regional-scale meteorological conditions and emissions rather than being confined to localized sources. This finding underscores the need for city-wide air pollution management strategies rather than isolated  interventions at specific locations.

\begin{figure}[H]
 \centering
  {\includegraphics[width=0.45\textwidth]{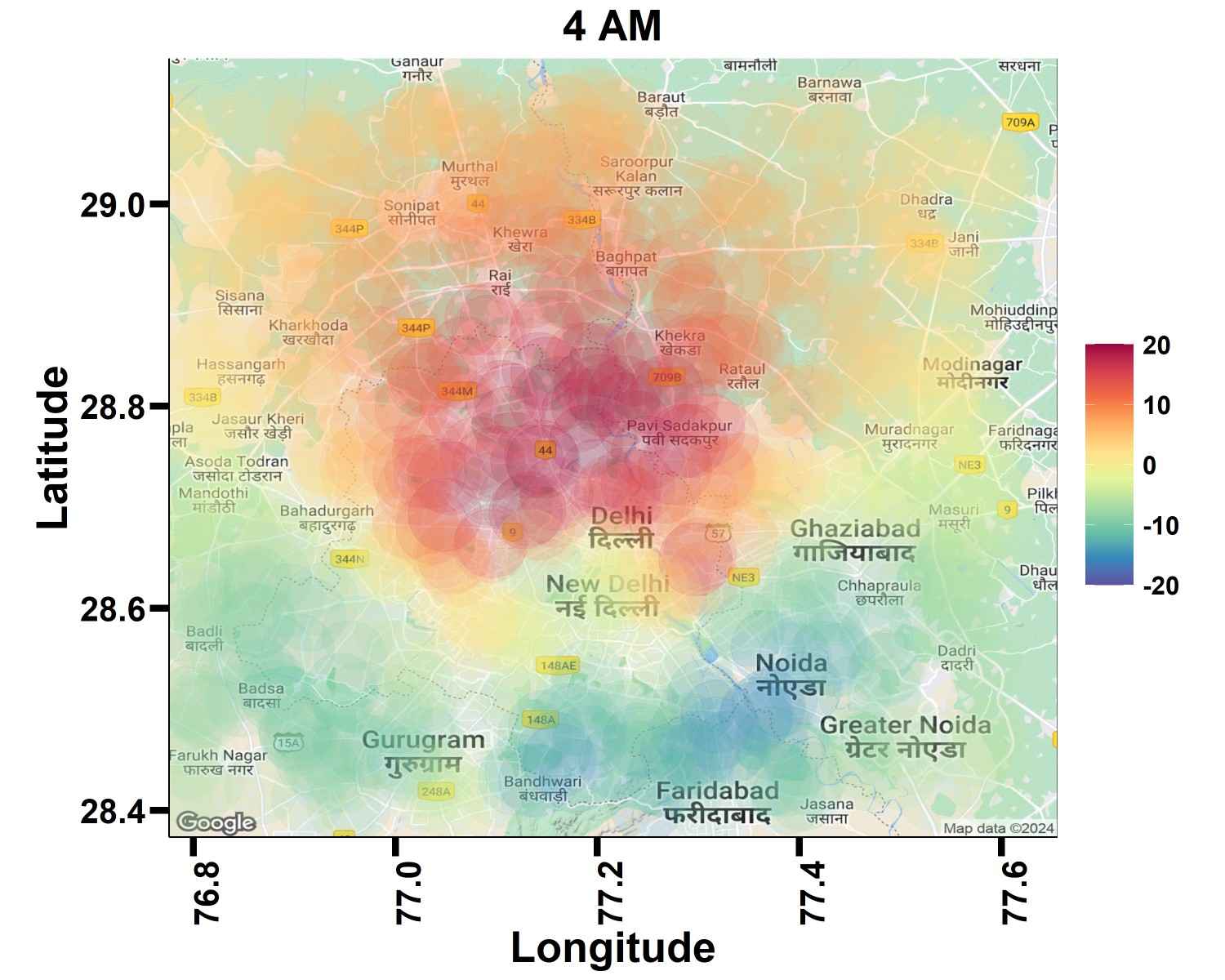}}
  {\includegraphics[width=0.45\textwidth]{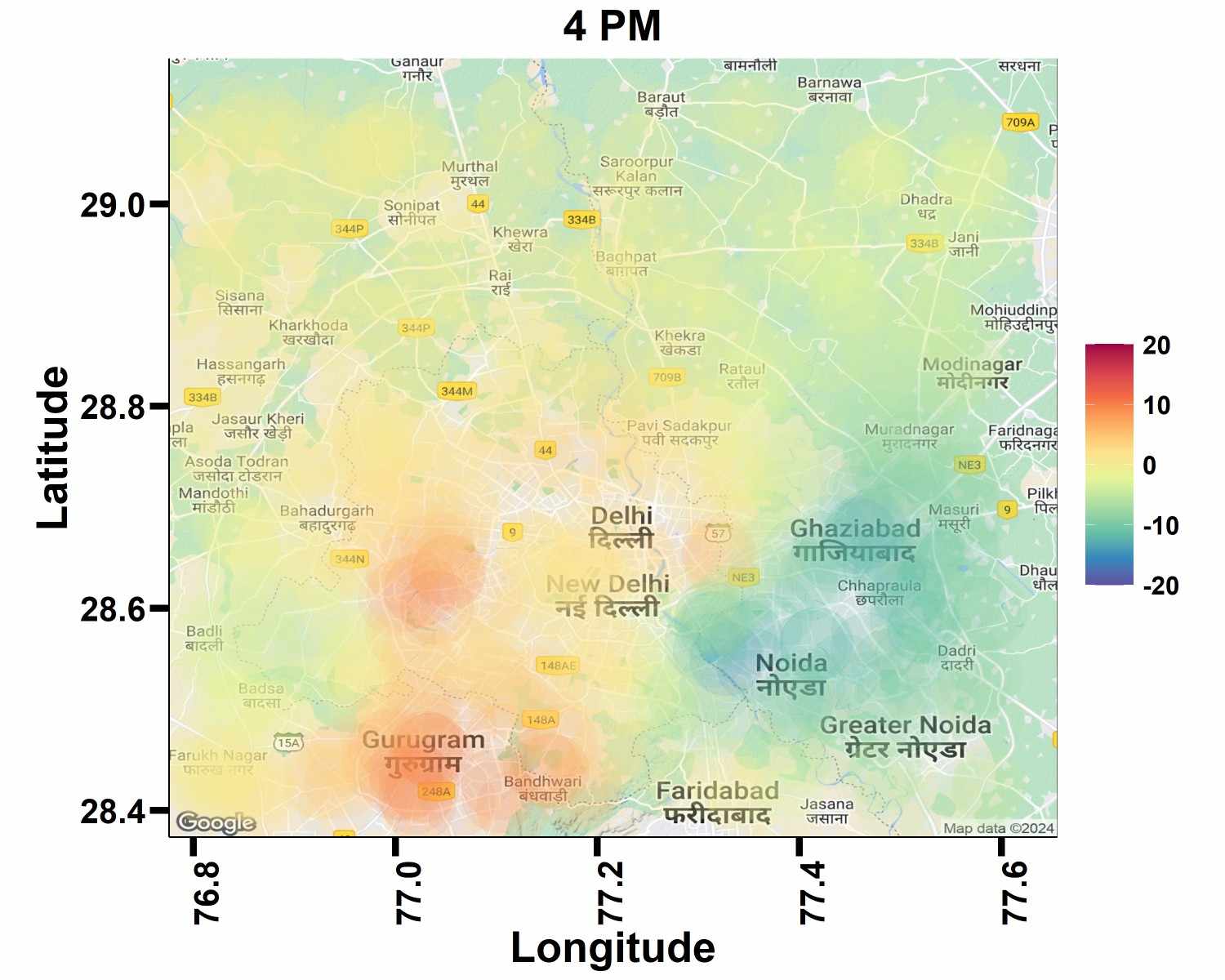}}
 \caption{Spatial distribution plots of detrended $PM_{2.5}$ levels in Delhi, generated using the Gaussian Process model (Equation~\ref{GP_correction}) on a 30$\times$30 grid with \texttt{R}. The maps reveal elevated $PM_{2.5}$ concentrations during early morning hours (4 AM), with significantly lower levels observed during the daytime (4 PM). This pattern is attributed to the expansion of the atmospheric boundary layer and enhanced vertical mixing during the day, which facilitate pollutant dispersion.
}
\label{fig_delhi_pm2.5_spatial_plot}
\end{figure}

To account for spatial correlation in air pollution data across different locations, we use a spatially correlated Gaussian process model\cite{Wang2023}. The estimated value \( \bar{y}_s(t) \) for the \( t^{\text{th}} \) hour at location \( s \) is given as:
\begin{equation}\label{GP_correction}
 \bar{y}_s(t) = \Sigma(s,s')[\Sigma(s,s') + \tau^2 \mathbf{I}]^{-1} y_s(t),
\end{equation}
where \( y_s(t) \) follows a Gaussian process with a mean of zero. The covariance function is defined as \( \Sigma(s,s') = \sigma^2 \exp(-\rho |s-s'|^2) \), \( \Sigma(s,s') \) models the spatial correlation between locations \( s \) and \( s' \), with \( \mathbb{V}ar(\epsilon) = \tau^2 \).

Figure~\ref{fig_delhi_pm2.5_spatial_plot} illustrates the spatial distribution of detrended $PM_{2.5}$ concentrations in Delhi at two representative hours; 4 AM and 4 PM; using a Gaussian Process regression model applied over a 30$\times$30 spatial grid in \texttt{R}. The early morning map (4 AM) shows higher $PM_{2.5}$ levels, particularly concentrated in the central and northern parts of the city, likely due to nocturnal accumulation of pollutants under stable atmospheric conditions and a shallow boundary layer. In contrast, the 4 PM map displays a marked reduction in $PM_{2.5}$ levels across most locations. This diurnal variation reflects the role of daytime boundary layer expansion and improved vertical mixing, which enhance pollutant dispersion. The contrast between these two time periods underscores the importance of atmospheric dynamics in shaping the spatial and temporal distribution of fine particulate matter in urban environments.

\subsection{Counterfactual cross-validation}

\subsubsection{Causal Impact of Emission Reductions on Air pollution in November}\label{sec:What-if-analysis}

Delhi experiences its highest levels of air pollution in November, particularly in terms of \( PM_{2.5} \) and \( O_3 \) concentrations, as revealed by Figures~\ref{fig:TS_Delhi} and \ref{fig_delhi_TS}. To estimate the causal effect of anthropogenic pollutants; \( NO_2 \), \( SO_2 \), and \( CO \); on \( PM_{2.5} \) and \( O_3 \), we fix the values of meteorological confounders, specifically atmospheric temperature (AT) and relative humidity (RH), at their median levels for November. This blocking of confounding pathways allows us to simulate a controlled intervention and isolate the treatment effect of pollutant reductions. The predictive Model~\ref{eqn_modl1} for \( PM_{2.5} \) and Model~\ref{modl_ozone} for \( O_3 \), introduced in Section~\ref{model:delhi_PM2.5}, are used to quantify the expected outcomes under different levels of the treatment variables.

The summary statistics of \( NO_2 \), \( SO_2 \), and \( CO \) concentrations, along with atmospheric temperature (AT) and relative humidity (RH) for November (2018 - 2023), are provided in Table~\ref{tab:summary_stat}. In this causal analysis, AT and RH are treated as confounding variables, as they influence both the concentrations of pollutants and the resulting levels of \( PM_{2.5} \) and \( O_3 \) through their effects on atmospheric dispersion and chemical transformation processes. To block these confounding pathways and isolate the treatment effects of \( NO_2 \), \( SO_2 \), and \( CO \), we fix AT and RH at their median November values; 21°C and 64\%; in both Model~\ref{eqn_modl1} and Model~\ref{modl_ozone}. This allows us to simulate a controlled intervention and estimate the causal impact of changes in pollutant levels on the air pollution outcomes.

The results of the causal scenario analysis, presented in Table~\ref{tab:scenario_1}, suggest a substantial treatment effect of \( NO_2 \), \( SO_2 \), and \( CO \) concentrations on \( PM_{2.5} \) levels. Keeping the meteorological confounders, atmospheric temperature, and relative humidity fixed at their median November values, we compared two treatment regimes: high versus low levels of anthropogenic pollutants. Under the high treatment condition (\( NO_2 \) = 110 \( \mu g/m^3 \), \( SO_2 \) = 25 \( \mu g/m^3 \), \( CO \) = 4.3 \( mg/m^3 \)), the predicted mean of \( PM_{2.5} \) is 308.32 \( \mu g/m^3 \), with a 95\% confidence interval of (154.20, 616.47). Under the low treatment condition (\( NO_2 \) = 28 \( \mu g/m^3 \), \( SO_2 \) = 8 \( \mu g/m^3 \), \( CO \) = 0.76 \( mg/m^3 \)), the predicted mean of \( PM_{2.5} \) reduces to 62.50 \( \mu g/m^3 \), with a 95\% confidence interval of (31.26, 124.97). The non-overlapping confidence intervals indicate that the estimated treatment effect is statistically significant, providing strong evidence that reductions in these pollutant levels causally reduce ambient \( PM_{2.5} \) concentrations.

In contrast, the response of \( O_3 \) to changes in pollutant levels exhibits an inverse but statistically insignificant effect. Under the high treatment condition; elevated levels of \( NO_2 \), \( SO_2 \), and \( CO \); the predicted mean concentration of ground-level \( O_3 \)  is 24.58 \( \mu g/m^3 \), with a 95\% confidence interval of (11.84, 51.00). When these pollutant levels are reduced to their lower bounds, the predicted mean \( O_3 \) increases to 45.04 \( \mu g/m^3 \), with a 95\% confidence interval of (21.70, 93.47). Although the point estimate for \( O_3 \)  increases under the low treatment condition, the overlapping confidence intervals suggest that this effect is not statistically significant. This behaviour is consistent with the well-established NO\textsubscript{x} titration mechanism, wherein high \( NO_2 \) levels suppress \( O_3 \)  formation through chemical scavenging\cite{li2019two}. Reducing \( NO_2 \) diminishes this scavenging effect, allowing for modest increases in \( O_3 \) concentrations. However, within the bounds of this analysis, the treatment effect on \( O_3 \)  remains statistically inconclusive.

Overall, these findings confirm that reducing anthropogenic emissions of \( NO_2 \), \( SO_2 \), and \( CO \) is an effective strategy for \emph{mitigating} fine particulate pollution in Delhi, as evidenced by the significant decline in \( PM_{2.5} \) levels. However, a holistic approach to air pollution management is necessary, as the interplay between pollutants can lead to unintended changes in other key air pollution indicators, such as \( O_3 \) . While the increase in \( O_3 \) is statistically insignificant in this scenario, further investigation into non-linear photochemical processes is warranted to ensure that \( O_3 \)  levels remain within safe limits when implementing emission control policies.

\begin{table}[H]
\begin{center}
\setlength{\arrayrulewidth}{0.7mm}
\setlength{\tabcolsep}{18pt}
\renewcommand{\arraystretch}{1.5}
\begin{tabular}{lccccc} \hline
\bf{Quantile} & \bf{NO$_2$} ($\mu g/m^3$) & \bf{SO$_2$} ($\mu g/m^3$) & \bf{CO} ($mg/m^3$) & \bf{AT} & \bf{RH}\\ \hline
\textbf{97.5\%} & 110 & 25 & 4.3 & 30 & 87\\  
\textbf{50\%} & 61 & 13 & 1.66 & 21 & 64\\  
\textbf{2.5\%}  & 28  & 8  & 0.76 & 15 & 32\\  
\hline
\end{tabular}
\caption{Summary statistics of \( NO_2 \), \( SO_2 \), and \( CO \) concentrations (\(\mu g/m^3\) and \( mg/m^3 \)), along with atmospheric temperature (AT, °C) and relative humidity (RH, \%) for November across the years 2018–2023. The table presents the 97.5\%, 50\% (median), and 2.5\% quantiles, highlighting the variability in air pollutant levels and meteorological conditions during this period.}
\label{tab:summary_stat}
\end{center}
\end{table}

\begin{table}[H]
\begin{center}
\setlength{\arrayrulewidth}{0.5mm}
\setlength{\tabcolsep}{10pt}
\renewcommand{\arraystretch}{1}
\small
\begin{tabular}{lccccc|cc} \hline
\bf{Quantile} & \bf{NO$_2$} ($\mu g/m^3$) & \bf{SO$_2$} ($\mu g/m^3$) & \bf{CO} ($mg/m^3$) & \bf{AT} (°C) & \bf{RH} (\%)& \bf{PM$_{2.5}$} & \bf{O$_3$}\\ \hline
\textbf{High Level} & 110 & 25 & 4.3 & 21 & 64 & 404.43 & 26.04\\ 
&&&&&& \small{(192.71~,~848.77)} & \small{(12.29~,~55.32)}\\ \hline
\textbf{Low Level}  & 28  & 8  & 0.76 & 21 & 64 & 68.25 & 43.98\\  
&&&&& & \small{(~32.52~,~143.23)} & \small{(20.73~,~93.31)}\\
\hline
\end{tabular}
\caption{This table presents a causal scenario analysis of predicted \( PM_{2.5} \) and \( O_3 \) concentrations under high and low levels of anthropogenic pollutants \( NO_2 \), \( SO_2 \), and \( CO \). To isolate the treatment effect of these pollutants, we block the influence of meteorological confounders; atmospheric temperature (AT, °C) and relative humidity (RH, \%); by fixing them at their median values for November. The table reports the expected mean values of \( PM_{2.5} \) and \( O_3 \), along with 95\% confidence intervals. The results support a statistically significant treatment effect on \( PM_{2.5} \), while the effect on \( O_3 \) is marginal and not statistically significant.}
\label{tab:scenario_1}
\end{center}
\end{table}

\subsubsection{Causal Inference Using the COVID-19 Lockdown as counterfactual validation}

To strengthen the causal claims made in Section~\ref{sec:What-if-analysis}, we now turn to real-world evidence from the COVID-19 lockdown in Delhi (March 25 - May 31, 2020), which serves as a natural counterfactual validation \cite{kment2020counterfactuals}. The lockdown, which sharply curtailed industrial activity and vehicular traffic, led to an exogenous reduction in anthropogenic emissions of \( NO_2 \), \( SO_2 \), and \( CO \). This unique setting allows us to empirically validate the estimated treatment effects on \( PM_{2.5} \) and \( O_3 \) by observing changes in pollution levels under actual emission reduction conditions. To implement this validation, we define the lockdown period in 2020 as the \textbf{treatment period} and compare it against analogous periods in 2019 and 2021, which serve as \textbf{Control Period 1} and \textbf{Control Period 2}, respectively. This quasi-experimental design enables us to assess the predictive power of our causal models and further verify that reductions in pollutant concentrations lead to meaningful improvements in air quality.

\begin{table}[H]
\begin{center}
\setlength{\arrayrulewidth}{0.5mm}
\setlength{\tabcolsep}{12pt}
\renewcommand{\arraystretch}{1.5}
\begin{tabular}{p{3.5cm}p{.5cm}p{.5cm}p{.5cm}p{.5cm}p{.5cm}p{.5cm}p{.4cm}}\hline
\bf{Period} & \bf{\( PM_{2.5} \) ($\mu g/m^3)$}  & \bf{O$_3$ ($\mu g/m^3$)} & \bf{NO$_2$ ($\mu g/m^3)$} & \bf{SO$_2$ ($\mu g/m^3)$} & \bf{CO\ $(mg/m^3)$} & \bf{AT\ $^\circ$C} & \bf{RH\ \%}  \\  \hline
\textbf{Control Period 1} & 85.05 & 48.26 & 52.28 & 21.26 & 1.31 & 31.14 & 32.09 \\ 
\small{March 25-May 31, 2019} &  & &  & &  &  & \\ \hline
\textbf{Treatment Period} & 48.37 & 52.33 & 22.98 & 14.90 & 0.77 & 29.67 & 45.95 \\ 
\small{March 25-May 31, 2020} &  & &  & &  &  & \\ \hline
\textbf{Control Period 2} & 69.93 & 45.61 & 35.40 & 15.73 & 1.01 & 29.28 & 41.67 \\  
\small{March 25-May 31, 2021} &  & &  & &  &  & \\ 
\hline
\end{tabular}
\caption{Average pollutant and meteorological values during the COVID-19 lockdown period in Delhi (March 25 - May 31, 2020), used as a natural counterfactual to validate the causal effects identified in Section~\ref{sec:What-if-analysis}. The table compares the \textbf{treatment period} (2020) with two \textbf{control periods} - the same timeframe in 2019 and 2021. Substantial reductions in anthropogenic pollutants (\( NO_2 \), \( SO_2 \), and \( CO \)) are observed during the lockdown, accompanied by a decrease in \( PM_{2.5} \) and a slight increase in \( O_3 \), consistent with the predicted treatment effects. Meteorological variables (AT and RH) are reported to control for confounding influences.}
\label{tab:what_if_analysis_2}
\end{center}
\end{table}

The COVID-19 lockdown served as a natural counterfactual experiment, providing an opportunity to assess the impact of reducing anthropogenic emissions on air pollution. Table~\ref{tab:what-if-analysis-Summary_stats} presents the observed and predicted levels of \( PM_{2.5} \) and \( O_3 \) during the lockdown (Treatment Period) in 2020, compared to the same timeframe in 2019 (Control Period 1) and 2021 (Control Period 2). The analysis focuses on April 27, with temperature fixed at 30°C and relative humidity at $40\%$ across all years to ensure consistent environmental conditions. Results indicate that during the lockdown, \( PM_{2.5} \) levels dropped significantly, with observed and predicted values of 42.55 \( \mu g/m^3 \) and 42.91 \( \mu g/m^3 \), respectively, compared to 87.86 \( \mu g/m^3 \) (2019) and 60.95 \( \mu g/m^3 \) (2021). The 95\% confidence interval (21.28, 85.07) was significantly lower than pre-lockdown values (43.94, 175.67), confirming a statistically significant reduction in fine particulate pollution. This decline aligns with the expected drop in transportation and industrial emissions, reinforcing the findings from our scenario analysis.  

In contrast, \( O_3 \)  levels exhibited an inverse response. The mean \( O_3 \) concentration increased from 48.26 \( \mu g/m^3 \) (2019) and 45.61 \( \mu g/m^3 \) (2021) to 52.33 \( \mu g/m^3 \) (observed) and 71.23 \( \mu g/m^3 \) (predicted) during the lockdown. However, the 95\% confidence intervals for all periods overlapped, indicating that the increase in \( O_3 \)  was statistically insignificant. This trend is consistent with the NO\textsubscript{x} titration effect, where reductions in \( NO_2 \) emissions lead to less \( O_3 \)  depletion at night, allowing ground-level \( O_3 \)  to accumulate. More importantly, the lockdown-induced reduction in emissions isolates the impact of human activities, allowing us to infer a \textbf{causal relationship} between reductions in \( NO_2 \), \( SO_2 \), and \( CO \) and the observed decrease in \( PM_{2.5} \). This aligns with the principles of causal inference, where a well-defined treatment (lockdown) and control framework provide strong evidence that anthropogenic pollution is a primary driver of fine particulate matter levels. While the observed \( O_3 \)  increase is minor and statistically insignificant, the findings highlight the complex chemical interplay between pollutants, underscoring the need for integrated air pollution management strategies that address both particulate matter and \( O_3 \)  precursors.  This analysis is consistent with analysis in Section \ref{sec:What-if-analysis}.


\begin{table}[H]
\begin{center}
\setlength{\arrayrulewidth}{0.5mm}
\setlength{\tabcolsep}{10pt}
\renewcommand{\arraystretch}{1.2}
\begin{tabular}{lccc} 
\hline
\bf{Summary Stats} & \bf{Control Period 1} & \bf{Treatment Period} & \bf{Control Period 2} \\
&March 25 - May 31, 2019&March 25 - May 31, 2020 & March 25 - May 31, 2021\\
\hline
\multicolumn{4}{l}{\bf{PM$_{2.5}$}} \\  
Mean from data & 85.04 & 48.37 & 69.93 \\  
Mean from model & 99.31 & 46.95 & 67.23 \\  
95\% CI for PM$_{2.5}$ & (47.32, 202.42) & (22.38, 98.58) & (32.03, 141.10) \\  
\hline
\multicolumn{4}{l}{\bf{O$_3$}} \\  
Mean from data & 48.26 & 52.33 & 45.62 \\  
Mean from model & 50.46 & 68.75 & 56.85 \\  
95\% CI for O$_3$ & (23.78~,~107.04) & (32.41~,~145.86) & (26.80~,~120.62) \\  
\hline
\end{tabular}

\caption{Predicted and observed \( PM_{2.5} \) and \( O_3 \) levels during the COVID-19 lockdown period (2020) compared to control periods (2019 and 2021). The analysis considers April 27 (mid-lockdown), with temperature fixed at 30°C and relative humidity at 40\% across all years. Mean values of \( NO_2 \), \( SO_2 \), and \( CO \) are derived from Table~\ref{tab:what_if_analysis_2}. Results indicate that reductions in \( NO_2 \), \( SO_2 \), and \( CO \) led to a significant decrease in \( PM_{2.5} \), while \( O_3 \) showed a marginal, statistically insignificant increase.}
\label{tab:what-if-analysis-Summary_stats}
\end{center}
\end{table}
 
\section{Summary}

This study presents a comprehensive analysis of the spatio-temporal dynamics and causal structure of air pollution in Delhi, focusing on the roles of key anthropogenic pollutants; \( NO_2 \), \( SO_2 \), and \( CO \); on two critical air pollution indicators: fine particulate matter (\( PM_{2.5} \)) and ground-level ozone (\( O_3 \)). Using a multifaceted approach that combines statistical modeling, time series decomposition, cross-correlation, Granger causality, and natural experiment validation, we investigate how these pollutants interact with meteorological variables and seasonal patterns to shape Delhi's air pollution over the period from January 2018 to August 2023. The descriptive and exploratory analysis reveals distinct diurnal and seasonal trends across pollutants. Elevated \( PM_{2.5} \) levels during night-time and early mornings are shown to be linked to shallow atmospheric boundary layers and reduced vertical mixing, while lower levels during the day correspond with solar-driven expansion of the boundary layer and enhanced dispersion. Ground-level \( O_3 \), on the other hand, follows a reverse pattern: increasing during daylight hours due to photochemical reactions involving \( NO_x \) and VOCs under sunlight, and decreasing at night due to titration by nitric oxide and surface deposition. These dynamics were further substantiated by the time series plots and scatterplots between \( PM_{2.5} \) and \( O_3 \), which showed a positive correlation during the day and a negative correlation at night.

To rigorously assess the impact of anthropogenic pollutants on air pollution, we developed a high-resolution regression model that includes meteorological variables (temperature and relative humidity), air pollutants, interaction terms, and periodic harmonics to capture seasonal and diurnal variations. This model achieves high explanatory power with an \( R^2 \) value of approximately 0.82 for \( PM_{2.5} \), and was extended to model \( O_3 \) levels as well. Multicollinearity in predictor variables was addressed via ridge regression, using 10-fold cross-validation to optimise the penalty term. The ridge-corrected models were used for all subsequent causal analysis to improve generalisability and interpretability of the coefficients. A central component of this study is the causal scenario analysis for the month of November, when Delhi experiences peak pollution levels. By fixing meteorological confounders at their median levels (21°C for temperature and 64\% for relative humidity), we isolate the treatment effect of changes in \( NO_2 \), \( SO_2 \), and \( CO \) concentrations on air pollution. The results show a statistically significant reduction in \( PM_{2.5} \) levels when pollutant concentrations are reduced from their upper to lower quantile values. The confidence intervals for predicted \( PM_{2.5} \) values under high and low treatments do not overlap, confirming the robustness of the causal effect. In contrast, although a marginal increase in \( O_3 \) levels was observed under low pollution scenarios, the overlapping confidence intervals indicate that this effect is not statistically significant. This behaviour aligns with the known NO\textsubscript{x} titration mechanism, which suppresses \( O_3 \)  accumulation under high \( NO_2 \) conditions.

To empirically validate these causal insights, we leveraged the COVID-19 lockdown in Delhi (March 25–May 31, 2020) as a natural counterfactual experiment. During this period, industrial activity and vehicular movement were drastically reduced, leading to significant declines in the concentrations of \( NO_2 \), \( SO_2 \), and \( CO \). We compared the lockdown period with the same timeframe in 2019 and 2021, treating them as control periods. The analysis revealed that both observed and predicted \( PM_{2.5} \) levels during the lockdown were substantially lower than in the control periods, while \( O_3 \) levels increased slightly but not significantly. These findings are consistent with the causal scenario analysis, further strengthening the evidence that reductions in anthropogenic emissions lead to improved air quality in terms of fine particulate pollution. However, it is important to acknowledge a limitation of using the COVID-19 lockdown as a counterfactual. The lockdown took place in April; part of the pre-monsoon summer season; when Delhi's pollution levels are typically lower due to stronger atmospheric mixing and fewer agricultural fires. As such, while the emission reductions during the lockdown are real, the extent to which they reflect the pollution dynamics of more critical months (e.g., November and December) is limited. This context means that the lockdown, although useful, should not be treated as a golden standard for causal inference but rather as a consistent validation point that must be interpreted with caution. The general agreement between the lockdown analysis and the November scenario analysis, however, offers compelling support for the direction and significance of our findings. In addition to these causal assessments, we examined cross-correlations between pollutants, revealing strong temporal linkages. Diurnal cyclic patterns emerged in the cross-correlation functions between \( PM_{2.5} \) and \( CO \), \( NO_2 \), and \( SO_2 \), as well as inverse relationships between \( O_3 \) and \( NO_2 \) or \( CO \). Granger causality tests further confirmed that \( NO_2 \) and \( CO \) are strong predictors of \( PM_{2.5} \), particularly during winter months, when meteorological conditions favour pollutant accumulation. The effects of \( SO_2 \) on both \( PM_{2.5} \) and \( O_3 \) were found to be weaker and more episodic, suggesting a less consistent role in Delhi's air pollution dynamics.

In conclusion, this study offers a robust assessment of air pollution dynamics in Delhi, combining descriptive analysis, statistical modeling, causal inference, and counterfactual validation. The findings underscore the significant role of anthropogenic emissions-- especially \( NO_2 \), \( CO \), and to a lesser extent \( SO_2 \) -- in driving fine particulate pollution, and suggest that emission control strategies targeting these pollutants could yield meaningful improvements in air quality. However, the complex interactions between primary and secondary pollutants, as well as the potential for unintended side effects on \( O_3 \)  levels, highlight the need for integrated and evidence-based air pollution management policies. While the COVID-19 lockdown provides valuable real-world validation, future studies should aim to extend causal inference approaches to other high-pollution periods and incorporate satellite-based and regional transport data for a more comprehensive understanding of pollution sources and dynamics in Delhi and beyond.

\section*{Acknowledgement}
The authors are grateful to Nalini Ravishanker, Ujjwal Kumar, and Amit Prakash for critical discussions.


\begin{thebibliography}{10}
\urlstyle{rm}
\expandafter\ifx\csname url\endcsname\relax
  \def\url#1{\texttt{#1}}\fi
\expandafter\ifx\csname urlprefix\endcsname\relax\def\urlprefix{URL }\fi
\expandafter\ifx\csname doiprefix\endcsname\relax\def\doiprefix{DOI: }\fi
\providecommand{\bibinfo}[2]{#2}
\providecommand{\eprint}[2][]{\url{#2}}

\bibitem{deBont2024}
\bibinfo{author}{de~Bont, J.} \emph{et~al.}
\newblock \bibinfo{journal}{\bibinfo{title}{Ambient air pollution and daily mortality in ten cities of india: a causal modelling study}}.
\newblock {\emph{\JournalTitle{The Lancet Planetary Health}}} \textbf{\bibinfo{volume}{8}}, \bibinfo{pages}{e433--e440}, \doiprefix\url{10.1016/S2542-5196(24)00114-1} (\bibinfo{year}{2024}).
\newblock \bibinfo{note}{Open access, published by Elsevier Ltd under CC BY 4.0}.

\bibitem{Dominici2022}
\bibinfo{author}{Dominici, F.} \emph{et~al.}
\newblock \bibinfo{journal}{\bibinfo{title}{Assessing adverse health effects of long-term exposure to low levels of ambient air pollution: Implementation of causal inference methods}}.
\newblock {\emph{\JournalTitle{Research Report Health Effects Institute}}} \textbf{\bibinfo{volume}{2022}}, \bibinfo{pages}{1--56} (\bibinfo{year}{2022}).

\bibitem{Brewer2023}
\bibinfo{author}{Brewer, D.}, \bibinfo{author}{Dench, D.} \& \bibinfo{author}{Taylor, L.~O.}
\newblock \bibinfo{journal}{\bibinfo{title}{Advances in causal inference at the intersection of air pollution and health outcomes}}.
\newblock {\emph{\JournalTitle{Annual Review of Resource Economics}}} \textbf{\bibinfo{volume}{15}}, \bibinfo{pages}{455--469}, \doiprefix\url{10.1146/annurev-resource-101722-081026} (\bibinfo{year}{2023}).
\newblock \bibinfo{note}{First published as a Review in Advance on June 20, 2023}.

\bibitem{ILENIC2024169117}
\bibinfo{author}{Ilenič, A.}, \bibinfo{author}{Pranjić, A.~M.}, \bibinfo{author}{Zupančič, N.}, \bibinfo{author}{Milačič, R.} \& \bibinfo{author}{Ščančar, J.}
\newblock \bibinfo{journal}{\bibinfo{title}{Fine particulate matter (pm2.5) exposure assessment among active daily commuters to induce behaviour change to reduce air pollution}}.
\newblock {\emph{\JournalTitle{Science of The Total Environment}}} \textbf{\bibinfo{volume}{912}}, \bibinfo{pages}{169117}, \doiprefix\url{https://doi.org/10.1016/j.scitotenv.2023.169117} (\bibinfo{year}{2024}).

\bibitem{SALONEN2019104887}
\bibinfo{author}{Salonen, H.}, \bibinfo{author}{Salthammer, T.} \& \bibinfo{author}{Morawska, L.}
\newblock \bibinfo{journal}{\bibinfo{title}{Human exposure to no2 in school and office indoor environments}}.
\newblock {\emph{\JournalTitle{Environment International}}} \textbf{\bibinfo{volume}{130}}, \bibinfo{pages}{104887}, \doiprefix\url{https://doi.org/10.1016/j.envint.2019.05.081} (\bibinfo{year}{2019}).

\bibitem{AKIMOTO1994213}
\bibinfo{author}{Akimoto, H.} \& \bibinfo{author}{Narita, H.}
\newblock \bibinfo{journal}{\bibinfo{title}{Distribution of so2, nox and co2 emissions from fuel combustion and industrial activities in asia with 1° × 1° resolution}}.
\newblock {\emph{\JournalTitle{Atmospheric Environment}}} \textbf{\bibinfo{volume}{28}}, \bibinfo{pages}{213--225}, \doiprefix\url{https://doi.org/10.1016/1352-2310(94)90096-5} (\bibinfo{year}{1994}).

\bibitem{Haiyan2019}
\bibinfo{author}{Haiyan~Qu, L.~L., Xiujun~Lu} \& \bibinfo{author}{Ye, Y.}
\newblock \bibinfo{journal}{\bibinfo{title}{Effects of traffic and urban parks on pm10 and pm2.5 mass concentrations}}.
\newblock {\emph{\JournalTitle{Energy Sources, Part A: Recovery, Utilization, and Environmental Effects}}} \textbf{\bibinfo{volume}{45}}, \bibinfo{pages}{5635--5647}, \doiprefix\url{10.1080/15567036.2019.1672833} (\bibinfo{year}{2019}).
\newblock \eprint{https://doi.org/10.1080/15567036.2019.1672833}.

\bibitem{Dmitri2022}
\bibinfo{author}{Kalashnikov, D.~A.}, \bibinfo{author}{Schnell, J.~L.}, \bibinfo{author}{Abatzoglou, J.~T.}, \bibinfo{author}{Swain, D.~L.} \& \bibinfo{author}{Singh, D.}
\newblock \bibinfo{journal}{\bibinfo{title}{Increasing co-occurrence of fine particulate matter and ground-level ozone extremes in the western united states}}.
\newblock {\emph{\JournalTitle{Science Advances}}} \textbf{\bibinfo{volume}{8}}, \bibinfo{pages}{eabi9386}, \doiprefix\url{10.1126/sciadv.abi9386} (\bibinfo{year}{2022}).
\newblock \eprint{https://www.science.org/doi/pdf/10.1126/sciadv.abi9386}.

\bibitem{urbansci7010009}
\bibinfo{author}{Yadav, R.~K.} \emph{et~al.}
\newblock \bibinfo{journal}{\bibinfo{title}{Relation between pm2.5 and o3 over different urban environmental regimes in india}}.
\newblock {\emph{\JournalTitle{Urban Science}}} \textbf{\bibinfo{volume}{7}}, \doiprefix\url{10.3390/urbansci7010009} (\bibinfo{year}{2023}).

\bibitem{zhang2022insights}
\bibinfo{author}{Zhang, J.} \emph{et~al.}
\newblock \bibinfo{journal}{\bibinfo{title}{Insights from ozone and particulate matter pollution control in new york city applied to beijing}}.
\newblock {\emph{\JournalTitle{npj Climate and Atmospheric Science}}} \textbf{\bibinfo{volume}{5}}, \bibinfo{pages}{85}, \doiprefix\url{https://doi.org/10.1038/s41612-022-00309-8} (\bibinfo{year}{2022}).

\bibitem{li2019two}
\bibinfo{author}{Li, K.} \emph{et~al.}
\newblock \bibinfo{journal}{\bibinfo{title}{A two-pollutant strategy for improving ozone and particulate air quality in china}}.
\newblock {\emph{\JournalTitle{Nature Geoscience}}} \textbf{\bibinfo{volume}{12}}, \bibinfo{pages}{906--910} (\bibinfo{year}{2019}).

\bibitem{ojha2022mechanisms}
\bibinfo{author}{Ojha, N.} \emph{et~al.}
\newblock \bibinfo{journal}{\bibinfo{title}{Mechanisms and pathways for coordinated control of fine particulate matter and ozone}}.
\newblock {\emph{\JournalTitle{Current pollution reports}}} \textbf{\bibinfo{volume}{8}}, \bibinfo{pages}{594--604} (\bibinfo{year}{2022}).

\bibitem{Geffner_etal_2022}
\bibinfo{editor}{Geffner, H.}, \bibinfo{editor}{Dechter, R.} \& \bibinfo{editor}{Halpern, J.~Y.} (eds.) \emph{\bibinfo{title}{Probabilistic and Causal Inference: The Works of Judea Pearl}}, vol.~\bibinfo{volume}{36} (\bibinfo{publisher}{Association for Computing Machinery}, \bibinfo{address}{New York, NY, USA}, \bibinfo{year}{2022}), \bibinfo{edition}{1} edn.

\bibitem{chakrabarti2023data}
\bibinfo{author}{Chakrabarti, A.~S.}, \bibinfo{author}{Bakar, K.~S.} \& \bibinfo{author}{Chakraborti, A.}
\newblock \emph{\bibinfo{title}{Data science for complex systems}} (\bibinfo{publisher}{Cambridge University Press}, \bibinfo{year}{2023}).

\bibitem{YADAV2024}
\bibinfo{author}{Yadav, A.}, \bibinfo{author}{Das, S.}, \bibinfo{author}{Bakar, K.~S.} \& \bibinfo{author}{Chakraborti, A.}
\newblock \bibinfo{journal}{\bibinfo{title}{Untangling climate’s complexity: Methodological insights}}.
\newblock {\emph{\JournalTitle{Indian Journal of Theoretical Physics}}} \textbf{\bibinfo{volume}{71}}, \bibinfo{pages}{9--29} (\bibinfo{year}{2024}).

\bibitem{Granger1969}
\bibinfo{author}{Granger, C. W.~J.}
\newblock \bibinfo{journal}{\bibinfo{title}{Investigating causal relations by econometric models and cross-spectral methods}}.
\newblock {\emph{\JournalTitle{Econometrica}}} \textbf{\bibinfo{volume}{37}}, \bibinfo{pages}{424--438} (\bibinfo{year}{1969}).

\bibitem{Wang2023}
\bibinfo{author}{Wang, J.}
\newblock \bibinfo{journal}{\bibinfo{title}{{ An Intuitive Tutorial to Gaussian Process Regression }}}.
\newblock {\emph{\JournalTitle{Computing in Science \& Engineering}}} \textbf{\bibinfo{volume}{25}}, \bibinfo{pages}{4--11}, \doiprefix\url{10.1109/MCSE.2023.3342149} (\bibinfo{year}{2023}).

\bibitem{kment2020counterfactuals}
\bibinfo{author}{Kment, B.}
\newblock \bibinfo{title}{Counterfactuals and causal reasoning}.
\newblock In \emph{\bibinfo{booktitle}{Perspectives on Causation: Selected Papers from the Jerusalem 2017 Workshop}}, \bibinfo{pages}{463--482} (\bibinfo{organization}{Springer}, \bibinfo{year}{2020}).

\end{thebibliography}

\end{document}